\documentclass[useAMS,usenatbib]{mn2e}
\usepackage{url,times,graphicx,amsmath,amsfonts,amssymb,aas_macros,color,epsfig,epstopdf}

\newcommand{\hMpc}{{\ifmmode{h^{-1}{\rm Mpc}}\else{$h^{-1}$Mpc}\fi}}
\newcommand{\hkpc}{{\ifmmode{h^{-1}{\rm kpc}}\else{$h^{-1}$kpc}\fi}}
\newcommand{\hMsun}{{\ifmmode{h^{-1}{\rm {M_{\odot}}}}\else{$h^{-1}{\rm{M_{\odot}}}$}\fi}}
\newcommand{\ltsima}{$\; \buildrel < \over \sim \;$}
\newcommand{\gtsima}{$\; \buildrel > \over \sim \;$}
\newcommand{\lsim}{\lower.5ex\hbox{\ltsima}}
\newcommand{\gsim}{\lower.5ex\hbox{\gtsima}}

\def\lesssim{\mathrel{\hbox{\rlap{\hbox{\lower4pt\hbox{$\sim$}}}\hbox{$<$}}}}
\def\gtrsim{\mathrel{\hbox{\rlap{\hbox{\lower4pt\hbox{$\sim$}}}\hbox{$>$}}}}

\newcommand{\beq}{\begin{equation}}
\newcommand{\eeq}{\end{equation}}
\def\beqa{\begin{eqnarray}}
\def\eeqa{\end{eqnarray}}
\def\hMpc{$h^{-1}\,{\rm Mpc}$}
\def\hkpc{$h^{-1}\,{\rm kpc}$}

\def\head{
 \vbox to 0pt{\vss
                   \hbox to 0pt{\hskip 440pt\rm LA-UR-10-07069\hss}
                  \vskip 25pt}}

\title[Baryon and thermal SZ properties in MUSIC]
{The MUSIC of Galaxy Clusters I: Baryon properties and  Scaling Relations of the thermal Sunyaev-Zel'dovich Effect}
\author[F. Sembolini et. al]
       {Federico Sembolini$^{1,2}$\thanks{E-mail: federico.sembolini@uam.es}, Gustavo Yepes$^{1}$, Marco De Petris$^2$, Stefan Gottl\"ober$^3$,\newauthor Luca Lamagna$^2$,  Barbara Comis $^2$\\
$^{1}$Departamento de F\'isica Te\'orica, M\'odulo C-15, Facultad de Ciencias, Universidad Aut\'onoma de Madrid, 28049 Cantoblanco, Madrid, Spain\\
$^2$Dipartimento di Fisica, Sapienza Universit\`a di Roma, Piazzale Aldo Moro 5, 00185 Roma, Italy\\
$^3$Leibniz-Institut f\"ur Astrophysik, An der Sternwarte 16, 14482 Potsdam, Germany
}

\begin{document}

\date{Accepted XXXX . Received XXXX; in original form XXXX}

\pagerange{\pageref{firstpage}--\pageref{lastpage}} \pubyear{2010}

\maketitle

\label{firstpage}

\begin{abstract}
We introduce  the Marenostrum-MultiDark SImulations  of galaxy Clusters (MUSIC)
 dataset.  It constitutes one of the largest sample of  hydrodynamically  simulated galaxy clusters with 
 more than 500 clusters and 2000 groups.   The objects have been selected from two large N-body simulations  and have been resimulated  at  high resolution using  Smoothed Particle Hydrodynamics (SPH)   together with  relevant physical processes that include cooling, UV photoionization, star formation and different feedback processes associated to Supernovae explosions.  

In this first paper  we focus on the analysis of the  baryon content
(gas and star) of  clusters in the MUSIC dataset  both as a function of
aperture radius and redshift.  The results from our
simulations are compared with  a compilation of the most recent observational
estimates of the gas fraction in  galaxy clusters at different
overdensity radii.  We confirm, as  in previous simulations, that the
gas fraction is overestimated  if radiative physics is not properly taken into
account. On the other hand, when the effects of
cooling and  stellar feedbacks are included,  the MUSIC clusters show a good
agreement with the  most recent observed  gas fractions  quoted in the
literature. A clear dependence of the gas fractions
with the  total cluster mass is also evident.  However, we do not find a significant
evolution with redshift  of the gas fractions at  aperture radius
corresponding to overdensities smaller than 1500  with respect to
critical density. At smaller radii, the gas fraction do exhibit a decrease
with redshift that is related  the gas depletion due to star formation in the central region of the clusters. 
The impact of the aperture radius choice, when comparing integrated quantities at different redshifts, is tested.  
The standard, widely used
definition of  radius at a fixed overdensity with respect to critical
density is compared with a definition of  aperture radius  based on the redshift dependent
overdensity with respect to background  matter density: we show that the latter
definition is more successful in probing the same fraction of the virial radius at different redshifts,  
providing a more reliable  derivation  of   the time evolution of
integrated quantities. 
We also present in this paper a detailed analysis of the scaling
relations of  the thermal SZ (Sunyaev Zel'dovich) Effect derived from MUSIC clusters.
The integrated SZ brightness, $Y$, is related to the cluster total
mass, $M$, as well as, the $M-Y$ counterpart which is more suitable for
observational applications. 
 Both laws are consistent with  predictions  from the self-similar
 model, showing a very low scatter which is $\sigma_{\log Y}$ $\simeq$ 0.04 and
 even a smaller one ($\sigma_{\log M}$ $\simeq$ 0.03) for the inverse $M-Y$
 relation.   The effects  of the gas fraction on the $Y-M$ scaling
 relation is also studied. At high  overdensities, the dispersion of
 the gas fractions  introduces non negligible deviation from
 self-similarity, which is directly  related to the $f_{gas}-M$
 relation. The presence of a possible redshift dependence on the $Y-M$
 scaling relation is also explored. No significant
 evolution of the SZ relations is found at  lower
 overdensities, regardless of the  definition of overdensity used.  

\end{abstract}
\noindent
\begin{keywords}
  methods: $N$-body simulations -- galaxies: clusters -- galaxies:
  scaling relations -- cosmology: theory -- Sunyaev-Zel'dovich effect
\end{keywords}

\section{Introduction} \label{sec:introduction}

Galaxy clusters are the biggest gravitationally bound objects of the
Universe and constitute one of the best cosmological probes  to measure the total matter content of the Universe. However, the total mass of these objects cannot be directly measured. It must be inferred from other observational quantities (X-ray  or SZ  Surface Brightness, Lensing distortions or number of galaxies). In all cases, one has to relate these quantities with the total mass of the system.  Due to the complex physics involved  in the processes of cluster formation,  hydrodynamical numerical simulations  have been a fundamental tool to   calibrate mass proxies, define  new ones, and to study the systematics affecting observational measurements.   They are  also  indispensable to  deeply study the
 formation and evolution of clusters of galaxies and all their gas-dynamical effects \citep{BK09}.

The big progresses achieved in the last years by numerical simulations
are well represented by their use to describe and to study X-ray
temperatures and their relation with cluster gas mass
(\citealt{ETTORI04}; \citealt{MUA06}; \citealt{NKV07})  as well  as to
compare numerical predictions  with observed temperatures (\citealt{LOKEN02};
\citealt{BORG04}; \citealt{LECCA08}) , gas profiles
(\citealt{RONCA06}; \citealt{CROS08}), or pressure profiles
\citep{ARNAUD10}. \textbf{A detailed review on cluster simulations can be found in \cite{BK09} and \cite{KB12}.}

In an  ideal scenario one would need   to have a large sample of  simulated galaxy clusters with   enough numerical resolution (both in mass and in the gravity and pressure forces) to  accurately resolve the internal substructures and with a detailed modelling of the most relevant physical processes. The best way to achieve this goal would be  by simulating large cosmological boxes (\citealt{BORG04}; \citealt{MN07}; \citealt{BURNS08}; \citealt{HAR08}; \citealt{BOY08}). Unfortunately, due to the large computational demand of  these  simulations, one needs to find a compromise  between the three main ingredients:  volume size,  mass resolution and physical processes included .  A possible  solution to  the computational  problems related with scalability of the present-day hydro codes is to proceed, mimicking  the observations, by  creating a catalogue of resimulated galaxy clusters  that are extracted from low resolution N-body simulations.  The  regions containing clusters of galaxies are then  resimulated  with  very high resolution, adding  only  gas physics  in the resimulated areas (\citealt{PUCH08};\citealt{DOLAG09}; \citealt{LAU09}; \citealt{FABJAN10}).  This so-called  'zooming' technique permits to simulate thousands of  clusters  basically independently from each other with less computational cost  than a full box  hydrodynamical simulation of the same resolution.  By selecting all the objects formed in a given volume above a given mass threshold,  mock  volume limited sample catalogues can be  generated and used in the study of the properties and  interrelations of the different scaling laws  of galaxy clusters.   Following this procedure,  we have generated  the MUSIC (Marenostrum-MultiDark SImulation of galaxy Clusters) dataset, a large sample of simulated clusters of galaxies, composed of objects extracted from two large box cosmological simulations: the MareNostrum Universe and the MultiDark simulation.
We selected all the clusters using criteria based on mass (selecting
all the clusters having a total mass larger than 10$^{15}$ \hMsun) or
on morphology (selecting groups of clusters corresponding to different
morphology classes, bullet-like clusters and relaxed clusters).  All
these objects have been resimulated with SPH particles, radiative physical  processes and star formation prescriptions, improving  by an order of magnitude the resolution with respect to the original
simulations, as described in the next section.  From this database, we will obtain mock observations  for  X-rays, SZ, lensing as well as optical galaxy counts. This will allow us to study the interrelations between the scaling laws associated to the different observables.   In this paper we  base our attention  on the properties of the  baryon content and the SZ effects and will leave a more detailed analyzes of   the relations with the other observed properties  for a further work. 

In the last few decades the  SZ  effect \citep{SZ70}  has become  one of the most powerful cosmological tool to study clusters of galaxies, as well as the nature of the  dark matter and dark energy components of the Universe.   The physical process  of the SZ effect  is the diffusion of CMB (Cosmic Microwave Background) photons with  a hot plasma  due to inverse Compton scattering.    The  thermal component of the SZ effect is largely enhanced  by the presence of clusters of galaxies, the most massive bound objects in the Universe, where plasma is in hydrostatic equilibrium inside the gravitational potentials  of dark matter. The Intra Cluster Medium (ICM) composed by high energy electrons constitutes an ideal laboratory to investigate the SZ effect. The brightness of the SZ  effect  turns out to be  independent of the diffuser position, thus making it  the best tool to find galaxy clusters at high redshift. Moreover, the  SZ flux collected from  the  cluster region  is proportional to the total thermal energy content, with a weak dependence on the  complex physical processes acting at  the inner regions (e.g. cooling flows,  galaxy feedbacks etc)   which mostly  affect  the X-ray luminosity.
Therefore, these two properties: redshift independence and low scatter mass proxy makes the  integrated Compton $Y$-parameter an efficient high-$z$ mass-estimator.

Under the hypothesis that the evolution of galaxy clusters is
driven mainly by gravitational processes \citep{KAI1986} and assuming hydrostatic equilibrium and an isothermal distribution of dark matter and ICM \citep{BN1998} it is possible to derive simple scaling power-laws linking cluster properties: the so-called self-similar scaling relations. In the case of SZ science, the relation linking the SZ brightness with the cluster total mass, the $Y-M$ scaling law, is continuously under analysis to test for its robustness, allowing the application of the SZ effect as a mass-finder.

Observational studies started to collect data of a few clusters,  mostly  those  with high X-ray luminosities (and therefore high masses)  (\citealt{BENS04}; \citealt{MORANDI07}; \citealt{BONA2008}; \citealt{VI09}; \citealt{ARNAUD10}). Recent large surveys have shown the possibility of detecting undiscovered clusters only through SZ effect observations, as in the claims by South Pole Telescope (SPT, \citealt{SPT2009}; \citealt{SPT2010}; \citep{SPT2011A}; \citealt{SPT2011B}), Atacama Cosmology Telescope (ACT, \citealt{ACT2011A}; \citealt{ACT2011B}) and Planck (\citealt{PLANCKa}; \citealt{PLANCKb}; \citealt{PLANCKc}; \citealt{PLANCKd}).
The possibilities to explore more distant objects or to deeply map
single cluster morphology are planned with the ongoing higher angular
resolution projects like AMI \citep{AMI08}, AMiBA \citep{AMIBA},
MUSTANG \citep{MUSTANG}, OCRA \citep{OCRA}, CARMA \citep{CARMA} or the
in-coming projects like the ground based C-CAT \citep{CCAT}, the
upgraded with new spectroscopic capabilities MITO \citep{MITO} and the balloon-borne OLIMPO \citep{OLIMPO}
or the proposed satellite mission
{\it Millimetron}  ({\ttfamily http:// www.sron.rug.nl/millimetron}).

More recent blind-surveys carried out by SPT, ACT and Planck enlarged
the existing dataset confirming the self-similarity in the sample at
least for the massive clusters (\citealt{ACT2011A}; \citealt{AND11};
\citealt{PLANCKc}).  Numerical simulations have shown that $Y$  (the
integral of the Compton $y$-parameter over the solid angle of the
cluster) is a good mass proxy (\citealt{DASILVA04}; \citealt{MOTL05};
\citealt{AGH09}) and that the slope and the evolution of SZ scaling
relations are apparently not affected by redshift evolution and cluster
physics. In order to find an X-ray equivalent of the SZ integrated
flux, the  numerical simulations also led to the introduction of a new mass
proxy: the $Y_X$ parameter, defined as the product of the cluster gas
mass and its tempearture \citep{YX06}.
 
Many works based on large $N$-body cosmological simulations have already studied the impact of gas physics on SZ scaling relations (\citealt{DASILVA00}; \citealt{HS02}; \citealt{DASILVA04}; \citealt{MOTL05}; \citealt{NAGAISZ}; \citealt{BONALDI07}; \citealt{HALL07}; \citealt{AGH09}; \citealt{BATTAGLIA11}; \citealt{KAY12}), such as the effects introduced by clusters with disturbed morphologies (\citealt{POOL07}; \citealt{WIK08}; \citealt{YANG10}; \citealt{KRAUSE12}), showing that the self-similar model is valid at least up to cluster scales (even if with some differences introduced by the different models used in the simulations). 

In what follows, we will present an analysis of  the relation between the SZ integrated flux and the total mass,  $Y$-$M$,  in MUSIC  clusters and will compare it  with the predictions of the  self-similar model.  We also study the possible biases introduced   by the common assumption of considering quantities ($M$ or $Y$) integrated inside a radius defined by a fixed  overdensity with respect to the critical density instead of a more suitable definition with respect to the  background density  whose value depends on redshift.

We will focus our  analysis of the $Y$-$M$ scaling relation on  the most massive objects of the MUSIC dataset which constitutes an  almost complete  volume limited sample. Therefore,  only clusters with virial masses $M_v>$5$\times$10$^{14}$\hMsun, are considered  in   this paper.  We will extend our analysis to  a wider range of masses in an upcoming work.

The paper is organized as follows. In Section 2 the MUSIC database  is described.  The baryon  content  in the clusters  at different  overdensities  is presented  in Section 3,  together with  a study of its  evolution with redshift and  a comparison of  our numerical results with the most recent  observational estimates. In Section 4 the $Y$-$M$ scaling relation is computed from  the MUSIC  clusters and compared with  the predictions from the self-similar model. We also discuss the validity of the integrated $Y$ as a proxy for the  cluster mass. In Section 5  we focus on  the impact of the gas fraction on the $Y$-$M$ relation  together with the dependence of the gas fraction on the cluster mass.
Finally in Section 6  the  redshift evolution  of  the $Y$-$M$  relation  is discussed.
In Section 7 we summarize and discuss our  main results.

\section{The MUSIC dataset} 
The MUSIC  project consists of two sets of resimulated clusters extracted from two large volume simulations:  
 
\begin{itemize}
\item  \emph{The MareNostrum Universe}, a non-radiative  SPH simulation with 2 billion particles (2$\times1024^3$ gas and dark matter )   in a 500$h^{-1}$ Mpc cubic box  \citep{MN07}.
\item \emph{The MultiDark Simulation}, a dark-matter only N-body simulation with $2048^3$ particles in a 1 $h^{-1}$Gpc cubic box  \citep{PETER11}.
\end{itemize}

These two simulations  have slightly different cosmologies. The Marenostrum Universe (MU) was made with  the cosmological parameters that were compatible with WMAP1 results 
( $\Omega_M$=0.3, $\Omega_b$=0.045, $\Omega_{\Lambda}$=0.7, $ \sigma_8$=0.9, $n$=1.0, $h$=0.7) 
while the MultiDark run was done  using the best-fit cosmological parameters to  WMAP7 + BAO + SNI ( $\Omega_M$=0.27, $\Omega_b$=0.0469, $\Omega_{\Lambda}$=0.73,  $\sigma_8$=0.82, $n$=0.95, $h$=0.7)  \citep{WMAP7}.

The procedure to  select  the interested objects in those two simulations  was also different.  For the MareNostrum clusters we mainly selected them based on the dynamical state (relaxed vs bullet-like clusters). For the MultiDark clusters  we  made a mass limited selection, taking all clusters with masses above  $10^{15}$ \hMsun   at  $z$ = 0. 

The zooming technique \citep{KLY01} was used to produce the  initial conditions of the resimulated objects. Namely, we first found all particles  within a sphere of 6 Mpc radius around the center of each selected object at $z$=0 from a low resolution version  ($256^3$ particles)  of the two simulations.  This set of particles was then mapped  back to the initial conditions to find out the Lagrangian region  corresponding to a  6 $h^{-1}$Mpc radius  sphere centred at the cluster centre of mass at $z$ = 0. The  initial conditions of the original simulations were  generated  in a finer mesh of $2048^3$  (for the MareNostrum) and $4096^3$  (for the MultiDark) sizes.  Therefore, we could improve the mass resolution of the resimulated objects by a factor of 8 with respect to the original  simulations.  We kept  the highest mass-refinement level  within the Lagrangian region of each   cluster and then cover it with shells of increasing mass particles  down to the lower resolution  level of $256^3$. To avoid problems with periodic conditions, we  always  recentre the simulations, each resimulated cluster always located at the center of the corresponding  box.  Thus, for the MultiDark clusters, we have dark matter particles of  5 different mass refinements ( from $4096^3$  to $256^3$) while for the MareNostrum clusters we have 4 different mass species. 
The gas SPH particles were added only to the highest refinement level. The SPH particle positions  were slightly displaced from their  parent dark matter by  0.4 times the mean inter particle distance in the 3 spatial directions and they were given the same initial velocity as their dark matter counterparts.  

The  parallel TreePM+SPH  GADGET code \citep{GAD05} was used to run all the resimulations.  We accounted for the effects of radiative cooling, UV photoionization, star formation and supernova feedback, including the effects of strong winds from supernova,  in the same way as described in \cite{STAR03} model. We refer  the reader to this reference  for a detailed description of the model.  Here we  simply provide the  values corresponding to the typical parameters  of the  model.  
Stars are assumed to form from cold gas clouds on a characteristic timescale $t_{\star}$ and a certain stellar mass fraction $\beta$ is instantaneously released due to supernovae from massive stars (M $>$ 8 $M_{\odot}$). We adopt a fraction of $\beta$=0.1   which is consistent  with assuming an Universal Salpeter IMF with a slope of $-1.35$  in the limits of 0.1 $M_{\odot}$ and 40 $M_{\odot}$. In addition to this mass injection of hot gas,  thermal  energy is also  released to the ISM by the supernovae. This feedback energy heats the ambient gas and cold clouds can be evaporated inside hot bubbles of supernovae. All these assumptions lead to a self-regulating star formation, where the growth of cold clouds is balanced by supernova feedback of the formed stars.  The number of collision-less star particles  spawned from  one SPH parent gas particle is fixed to 2 , which means that each stellar particle will get half of the original parent SPH mass.   In addition to the thermal feedback,   some sort of kinetic feedback is also accounted for.  Gas mass losses due to galactic winds, $\dot{M}_{\rm w}$, is assumed to be proportional to the star formation rate  $M_{\rm SFR}$, i.e. $\dot{M}_{\rm w}$=$\eta M_{\rm SFR}$ with $\eta$=2. Additionally, the wind contains a fixed fraction $\epsilon$ of the total supernova energy, which is assumed to be $\epsilon=0.5$.  SPH particles near the star formation region will be subject to enter in the wind   in an stochastic way. Those particles eligible for winds will be given an isotropic velocity kick  
of $v$=400 km/s and will freely travel without feeling pressure forces up to 20 kpc  distance from their  original positions.

Using the aforementioned  techniques  we have been able to resimulate a total of 700 different galaxy clusters from the two large box simulations.  This constitutes one of the largest samples of resimulated cluster-sized  objects.   We used two different selection criteria, based on dynamical state and on mass cuts  in order to explore scaling relations dependencies  on  different aspects (mass, morphology, redshift evolution, overdensity).
  All these clusters were also resimulated with non-radiative physics,
  using only SPH and gravity forces, so we can also account for the
  effects of cooling,  star formation and  supernovae feedbacks on the
  observed properties of the clusters. 
The  database of MUSIC clusters is publicly available in a SQL query
format through the website:  {\ttfamily http://music.ft.uam.es}. The
database will comprehend also the initial conditions of all MUSIC
objects, in order to give the possibility to resimulate the clusters
with other  hydro codes and/or with different  modelling for the
radiative processes (e.g. AGN feedback, MHD, Cosmic ray pressure, etc).
 
In what follows we describe in more detail the two  main subsamples in which the MUSIC database is divided.


\subsection{MUSIC-1: MareNostrum resimulated clusters}\label{sec:results}
This  first subset is composed of 164 resimulated clusters extracted from the MU.
 
There are  more than 4000 clusters of galaxies (objects larger than 10$^{14}$ \hMsun) in the whole MU simulation.
Out of  this large number of clusters, we selected 82 'bullet-like' clusters. The 1ES0657-556 cluster \citep{CLOWE06}, generally known as the bullet cluster, is considered one of the best astrophysical tools to study the nature of dark matter. This cluster, located at $z= 0.296$, is composed by a smaller merger which has just crossed a bigger object and is now moving away. The core of the plasma cloud of the merging sub-cluster, 'the bullet', has survived the passage. Due to the different dynamics of the collision less dark matter component and the x-ray emitting gas,  it is expected  that   the two  components have a spatial offset.   By combining lensing and X-ray observations, this behaviour is seen explicitly  in the bullet cluster.   The relative velocity of the two merging  systems   is a quantity that is not directly measured. It is derived from the shape of the "bow  shock"  clearly seen in  X-ray images,  using the results from different  hydro simulations of merger clusters (\citealt{SPRING07}; \citealt{Mastro08}). But a more direct  measure of   the effects of a cluster merger, in the  gas and dark matter components,  is the projected 2D separation between the peak of the two distributions (gas and  dark matter).  \cite{BULLET10} have used this measurement to characterize the number of bullet-like clusters in the MU.   They define a "bullet" cluster if the 2D separation between the peak of dark matter and the peak of gas distribution is at least 200 kpc (as in the real bullet cluster). 

In order to have a morphological counterpart to  these extremely disturbed systems, we also selected  another 82 clusters   which exhibit  the following  relaxed conditions:  the  displacement between dark and baryonic matter  is smaller than 200 kpc  and  they are composed of a single massive cluster  with no big  substructures  (i.e.  all substructures inside the virial radius  must have masses  smaller than 10 per cent of  the total mass).

 The relaxed clusters were chosen  to have masses similar to  those of the bullet systems: in this way every bullet-like cluster has  a relaxed companion of  the  same mass. All the selected clusters were resimulated using the zooming technique and radiative physics, including radiative cooling, heating processes of the gas arisen from a UV background, star formation and supernovae feedback explained above. The particle mass  of dark matter was set to m$_{DM}$=1.03$\times$10$^9$\hMsun and the mass of  SPH  gas particle to m$_{gas}$=1.82$\times$10$^8$\hMsun. Due to the relative small size of the computational box of the MU simulation,   very few clusters of MUSIC-1 subset have a mass at the virial radius bigger than 10$^{15}$ \hMsun. The cluster masses  of MUSIC-1 span from  \emph{M$_{v}$}=2$\times$10$^{15}$\hMsun  to \emph{M$_{v}$}=1$\times$10$^{14}$\hMsun.

\subsection{MUSIC-2 : MultiDark resimulated clusters} \label{sec:results}
The second, and more numerous, subset of MUSIC consists of  a mass limited sample of resimulated clusters selected from the MultiDark Simulation (MD).  This  simulation is  dark-matter only and  contains 2048$^3$ (almost 9 billion) particles in a (1$h^{-1}$Gpc)$^3$ cube. It was performed in 2010 using ART \citep{ART} at the NASA Ames Research centre. All the data of this simulation are accessible from the online {\itshape MultiDark Database} (www.MultiDark.org). We have selected, using  a  low resolution (256$^3$ particles) version of the MD simulation,  all the objects more massive than 10$^{15}$ {\itshape  \hMsun}. 
In total,  we found 282 objects above this mass limit. All these massive clusters were resimulated  both with and without radiative physics using the techniques explained above.  The mass resolution for these simulations corresponds  to m$_{DM}$=9.01$\times$10$^8$\hMsun and  to m$_{SPH}$=1.9$\times$10$^8$\hMsun.  The gravitational softening was set to 6 $h^{-1}$ kpc for the SPH and dark matter particles in the high-resolution areas. 
Several low mass clusters have been found close to the large ones and
not overlapping with them. Thus, the total number of resimulated
objects is considerably larger.  In total we obtained  535  clusters
with M$>$10$^{14}$ \hMsun at $z$ = 0 and more than 2000 group-like objects
with masses in  the range 10$^{13}$ \hMsun$<$M$_v<$10$^{14}$ \hMsun.
All of these resimulations have been done both with non-radiative
physics, using the same technique and resolutions as the radiative
ones.  

\begin{figure}
\centering\includegraphics[angle=0,width=8cm]{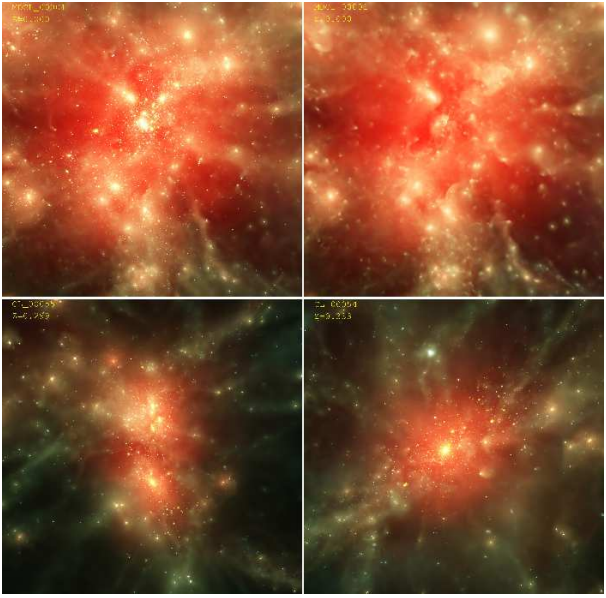}
\caption{Gas distribution in MUSIC clusters: the physics at $z$ = 0; a bullet-like cluster(bottom-left) and a relaxed cluster (bottom-right) of MUSIC-1 at $z$ = 0.3. The images of the all MUSIC dataset (available at {\ttfamily http://music.ft.uam.es}) have been generated with SPLOTCH \citep{SPLOTCH}. }
\label{Castor}
\end{figure}

We have  stored snapshots for  15 different redshifts in the range 0 $\leq$ 
$z$ $\leq$ 9 for each resimulated object. In the case of the two massive
clusters, we stored 320 snapshots for a more detailed study of the
formation process.  A graphical representation of the  gas distribution
for all the MUSIC-2 clusters for each of the 15 redshifts in radiative
and non-radiative runs can be seen online at  {\ttfamily
  http://music.ft.uam.es}. In Fig.\ref{Castor} we show, as an
example, the gas density, colour coded according to the temperature,
and the stellar component for  4 different clusters of the MUSIC
database at $z = 0$.

\begin{figure}
\centering\includegraphics[angle=0,width=9cm]{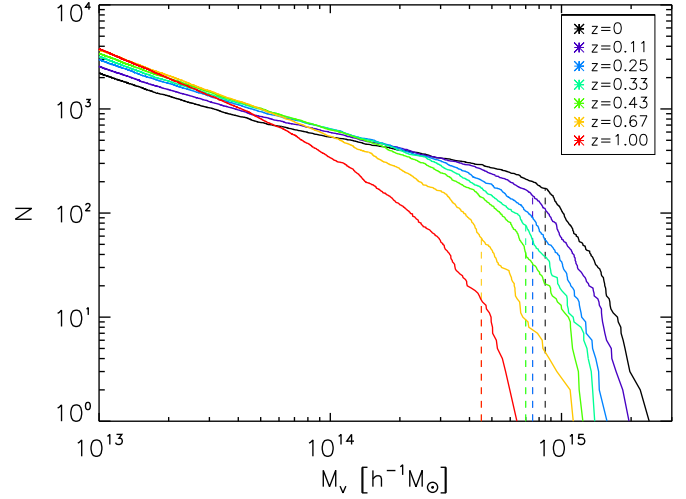}
\caption{Cumulative mass function of the MUSIC-2 dataset from $z$ = 0 to $z$ = 1. The vertical lines indicate the mass limit beyond which the dataset constitutes a complete volume limited sample}
\label{Mfunc}
\end{figure}

The cumulative mass function of MUSIC-2 clusters in the redshift range
0 $\leq$ $z$ $\leq$ 1 is shown in figure \ref{Mfunc}. The mass values of the dataset
completeness for all redshifts analyzed are shown in Table
\ref{Mcompl}, spanning from 4.5$\times$10$^{14}$ \hMsun at $z$=1 to
8.5$\times$10$^{14}$ \hMsun at $z$ = 0.

In this first paper about MUSIC clusters, we focused our analysis only
on the most massive objects of MUSIC-2, selecting all the clusters of
the dataset, both  radiative (CSF) and non-radiative (NR), with 
virial mass $M_v>$5$\times$10$^{14}$ \hMsun at $z = 0$. This selection allows us
to analyze a subset in a mass range where MUSIC constitutes a complete 
volume limited sample  (more than 80 per cent of the
clusters found in the 1 $h^{-1}$Gpc MD box above this mass limit have
been resimulated by MUSIC). Moreover,  those clusters  are  the best resolved ones , containing millions of particles. This  allows us to to extend the analysis towards the inner regions of the
clusters. The selected subset has been analyzed at seven different
redshifts. In   Table  \ref{Mcompl},  we show the actual redshifts analyzed and the total number of clusters used in each redshift bin. We did not include any of the MUSIC-1 clusters in this analysis, as only very few of them have a virial mass larger than 5$\times$10$^{14}$ \hMsun at $z = 0$. \textbf{All these massive objects are free from contamination of low resolution particles. The distance between the center of the cluster and the closest low resolution particle is at least 2 times the virial radius at $z$ = 0 and more than 3 times the virial radius at higher redshifts.}

\begin{table}
\begin{center}
\begin{tabular}{|c|c|c|c|c|c|c|c|}
\hline
$\textbf{z}$ & 0.00 & 0.11 & 0.25 & 0.33 & 0.43 & 0.67 & 1.00 \\
\hline
\textbf{M$_v$} &  8.5 & 7.5 & 7.5 & 7.0 & 7.0 & 
4.5 & 4.5 \\
\hline
$\textbf{N}$ & 172 & 147 & 90 & 76 & 41 & 57 & 14 \\
\hline
\end{tabular}
\end{center}
\caption{Mass completeness of MUSIC-2 dataset (compared to dark-matter only MD simulation) in the redshift range 0 $\leq$ $z$ $\leq$1. The virial mass is reported in 10$^{14}$ $h^{-1}M_{\odot}$ and $N$ is the number of clusters beyond the completeness mass limit.}
\label{Mcompl}
\end{table}


\subsection{Halo-finding and general properties of the clusters}
Haloes and sub-haloes of MUSIC simulations have been identified and their properties have been measured using the hybrid MPI+OpenMP parallel halo finder AHF (see \citealt{KK09} for a detailed description), a code whose algorithm automatically
identifies haloes, sub-haloes, sub-subhaloes, etc. 

 It has been proven that it can reliably recover
substructures containing at least 20 particles \citep{HALO11}.

The luminosity of the stars formed in the simulation has been computed
using the STARDUST \citep{DEV99} code, a stellar population synthesis
model of the spectral energy distributions (SEDs) of star bursts from
far$-$UV to radio wavelengths. From the SEDs of the stars we could
compute the total luminosity in different photometric bands for each
object of the simulation.

Simulated X-rays maps of our clusters will be produced using X-MAS
(\citealt{GARD04}; \citealt{RASIA08}) and PHOX \citep{PHOX}, tools
developed in order to simulate X-ray observations of galaxy clusters
obtained from hydrodynamical $N$-body simulations. One of the main
characteristics of these codes  is to generate event files following the
same standards used for real observations, allowing to analyze
simulated observations with the same tools of real observations.

\begin{figure}
\centering
\includegraphics[angle=0,width=9cm]{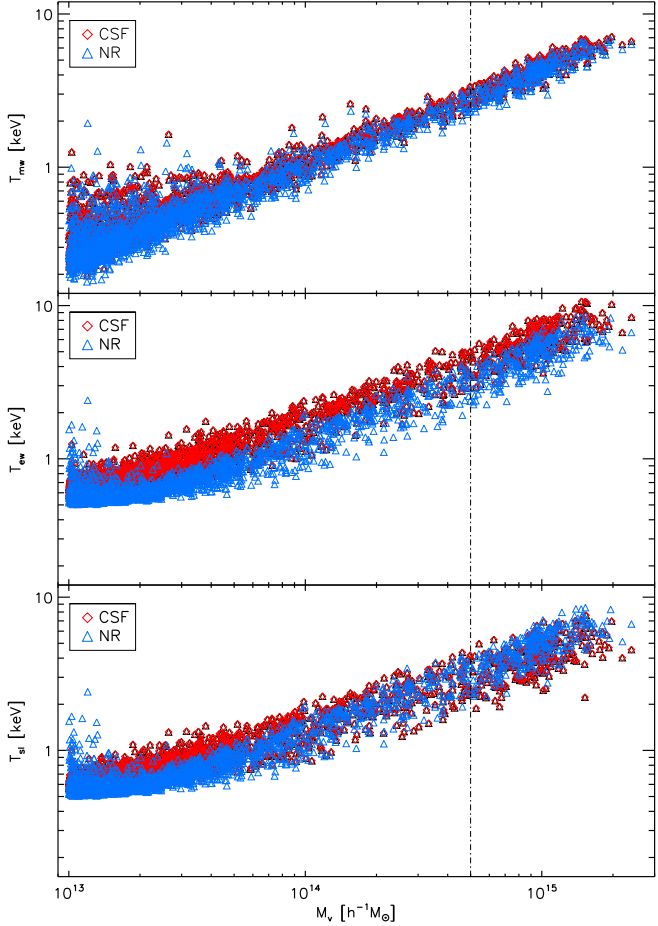}
\caption{Distribution of mass-weighted (top panel), emission-weighted (middle panel) and spectroscopic-like (bottom panel) temperatures of all MUSIC-2 objects (clusters and groups) at virial radius as a  function of the virial  mass at z = 0. Red diamonds refer to CSF objects, blue triangles to NR objects. The dashed vertical line indicates the subset analyzed in this paper ($M_v>$5$\times$10$^{14}$ \hMsun at $z$ = 0). }
\label{T}
\end{figure}

Here,  we provide  the  estimated  X-ray temperatures of the simulated clusters as:
\begin{eqnarray}
&&T_{mw} = \frac{\sum_i T_im_i}{\sum_im_i}\label{mw}\\
&&T_{ew} =  \frac{\sum_i m_i \rho_i \Lambda(T_i)T_i}{\sum_i m_i \rho_i \Lambda(T_i)}\label{ew}\\
&&T_{sl} = \frac{\sum_i \rho_i m_i T_i^{1/4}}{\sum_i \rho_i m_i T_i^{-3/4}} \label{sl}
\end{eqnarray}
calculating the mass weighted temperature (Eq.\ref{mw}), the emission weighted temperature (Eq.\ref{ew}) and the spectroscopic-like temperature (Eq.\ref{sl}) \citep{MAZ04}, where $m_i$, $\rho_i$, $T_i$ are the gas particle mass, density and electronic temperature and $\Lambda(T)$ is the cooling function. The particle gas densities of the CSF runs have been corrected in order to take into account the multiphase model adopted in the radiative simulation; the spectroscopic-like and emission-weighted temperatures have been calculated only by considering gas particles with $kT>$0.5 keV and we have assumed 
$\Lambda(T) \propto  \sqrt{T}$ corresponding to thermal bremsstralung.
 For all the MUSIC-2 objects (clusters and groups) the three  temperature definitions, estimated at the virial radius, are shown in Fig. \ref{T} along with  the corresponding virial mass, $M_v$, at $z$ = 0 for both the simulations (NR and CSF). We will have a more detailed analysis of the cluster X-ray properties in a following papers.

\section{Baryon content  of  MUSIC-2 massive clusters}\label{sec:bar}
We study the internal baryon content of the MUSIC-2 cluster sample
estimating the gas and stellar  fractions, as well as the total baryon
fraction. In this way we are able to check the consistency between our simulated 
dataset and available observational results. Moreover, as examined in
the following section, the gas fraction plays a relevant role on the
$Y-M$ scaling relation. The over cooling effect, a hot topic in all
cluster simulations \citep{KRAV05}, is also monitored in this dataset.

The mass of each component is estimated inside  fixed overdensity radii  defined either with respect  to the critical or background density. In both cases, we calculate the gas mass $M_{gas,\Delta}$, the star mass $M_{star,\Delta}$ and the total mass $M_{\Delta}$ inside a fixed overdensity radius $r_{\Delta}$ selecting and summing all the particles of gas, star and dark matter falling inside that radius.   We then define the  baryon, gas and star fractions as:
\begin{eqnarray}
&&f_{gas}=\frac{M_{gas,\Delta}}{M_{\Delta}}\\
&&f_{star}=\frac{M_{star,\Delta}}{M_{\Delta}}\\
&&f_{bar}=\frac{M_{gas,\Delta}+M_{star,\Delta}}{M_{\Delta}}
\end{eqnarray}

These quantities are calculated  for the CSF clusters. In the case
of the NR simulations, where all the SPH particles are gas, only the
baryon fraction,  defined as $f_{bar}$=$M_{gas,\Delta}$/$M_{\Delta}$ is
estimated.

From now on,  we are going to compare  the integrated cluster properties  at different redshifts. It  is therefore very important  the definition of the aperture radius where the  properties are estimated. The most often used, more favourable in terms of simplicity, consists on  defining  the  radius  at which the  mean enclosed total  density is a fixed  factor (independent on redshift)  of  the critical density of the Universe, $\rho_c$($z$). This is useful to make comparisons with theoretical predictions from the spherical collapse model, which describe the virialized part of a cluster in terms of  the density contrast $\Delta_v$. But  $\Delta_v$ varies with redshift and depends on the cosmological parameters adopted. 
Thus, as we show in Appendix A, this definition of aperture radius based on the same  critical overdensity value  at different redshift  does not allow to consider  a region of  the clusters corresponding to the same fraction  of the virial radius, $R_v(z)$ at different redshift.  This could introduce a bias  in the study of the redshift evolution of  integrated cluster quantities simply due to the different  cluster regions  probed at  different redshifts. To alleviate this problem, it is useful to adopt another definition of aperture radius   based on the redshift-dependent background density, $\rho_b$($z$).  This approach was introduced   the first time by \cite{MGAN06} to study the evolution of cluster X-ray scaling relations, but it has never been considered in the study of the  SZ relations before.  In this paper we use both definitions of aperture radii:  a fixed value, independent on redshift,  critical overdensity and a redshift-dependent background overdensity.  We refer the reader to Appendix A in which we compare the   redshift evolution of the radii for these two definitions. 

In order to have a detailed description of the integrated cluster properties starting from the inner regions up to the virial radius, we analyze our clusters at seven different critical overdensities, $\Delta_c$=[$\Delta_{v,c}$($z$), 200, 500, 1000, 1500, 2000, 2500], where only the virial value is different for each redshift as in Eq.(\ref{vir2}). 
Similarly we select a set of (redshift dependent) background overdensities, $\Delta_b$($z$)=[$\Delta_{v,b}$($z$), 500($z$), 1000($z$), 1500($z$), 2500($z$), 5000($z$), 7000($z$)], where we are  adopting the notation used by  \cite{MGAN06}.  We are referring to an overdensity $\Delta_b$($z$)  corresponding to the value at $z$ = 0 but that is changing its value along the redshift (i.e. writing 7000($z$)  it means that we  assume  $\Delta_b$=7000 at $z$ = 0 and varying the value with redshift as in Eq.(\ref{vir})). These  values have been chosen aiming at  having the same fraction  of the clusters volume  than  those analyzed with a fixed critical overdensity criteria.
 In this way we can infer the properties at overdensities
 corresponding to cluster regions that are well fitted by X-ray observations or  by
 current SZ observations.  Due to the high resolution of our simulations, we
 can extend the analysis down  to  the cluster innermost regions
 ($\Delta_c$ = 2500). Overdensities higher than 2500, corresponding to the
 core region of the cluster, can be affected in a non negligible way by
 cold flows or  galaxy feedbacks.

\begin{figure}
\centering\includegraphics[angle=0,width=9cm]{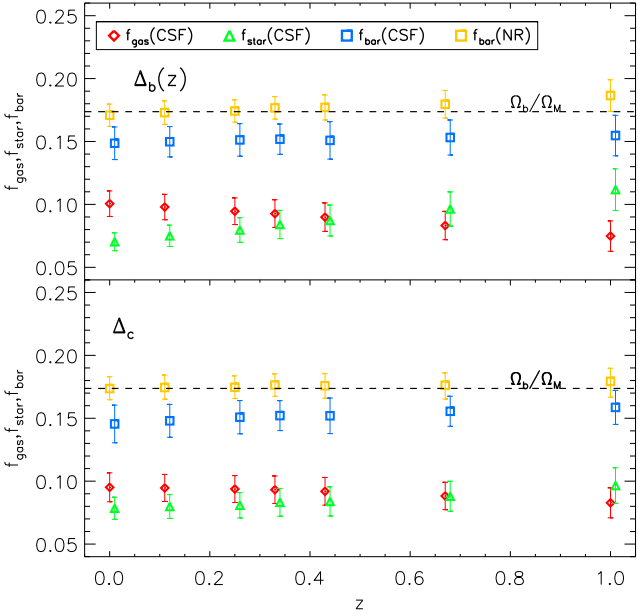}
\caption{Evolution of $f_{gas}$ (red diamonds), $f_{star}$ (green triangles) and $f_{bar}$ (yellow squares) of CSF clusters and $f_{bar}$ (blue squares) of NR clusters for $\Delta _b$=7000($z$) (top panel) and $\Delta _c$=2500 (bottom panel).}
\label{f2500}
\end{figure}

Fig.\ref{f2500} shows the evolution of the gas, star and baryonic fractions
in a  region close to the cluster centre, ($\Delta_b$=7000($z$) and $\Delta_c$=2500), corresponding at almost $R_v$/5. When we consider $\Delta_c$=2500 we
observe a moderate decrease in the gas fraction  as we go from $z$ = 0
($f_{gas}$=0.09$\pm$0.01) to $z$ = 1 ($f_{gas}$=0.08$\pm$0.01). Correspondingly, 
the value of the star fraction grows from $f_{star}$=0.08$\pm$0.01
at $z$ = 0 to $f_{star}$=0.10$\pm$0.01 at $z$ = 1. The baryon fraction shows, however, 
no significant variation with redshift, and it is compatible with the
cosmic value $\Omega_b/\Omega_m$=0.174 \citep{WMAP7} for CSF clusters
($f_{bar}$=0.174$\pm$0.009), while it is smaller for NR clusters
($f_{bar}$=0.15$\pm$0.01). We find a similar scenario when we define the aperture radius based on  $\Delta_b$=7000($z$), but in this case the evolution of gas and star
fraction with redshift is more evident: from $f_{gas}$=0.09$\pm$0.01 and
$f_{star}$=0.07$\pm$0.01 at $z$ = 0 to $f_{gas}$=0.07$\pm$0.01 and
$f_{star}$=0.11$\pm$0.02 at $z$ = 1. We observe that at these overdensities
the value of the star fraction is high, especially at high redshifts, but
the effects of a possible over cooling  in our simulations do not seem to be 
strong  enough  to deviate the mean values of the gas fraction from  the  observational  estimates. 

\begin{figure}
\centering\includegraphics[angle=0,width=9cm]{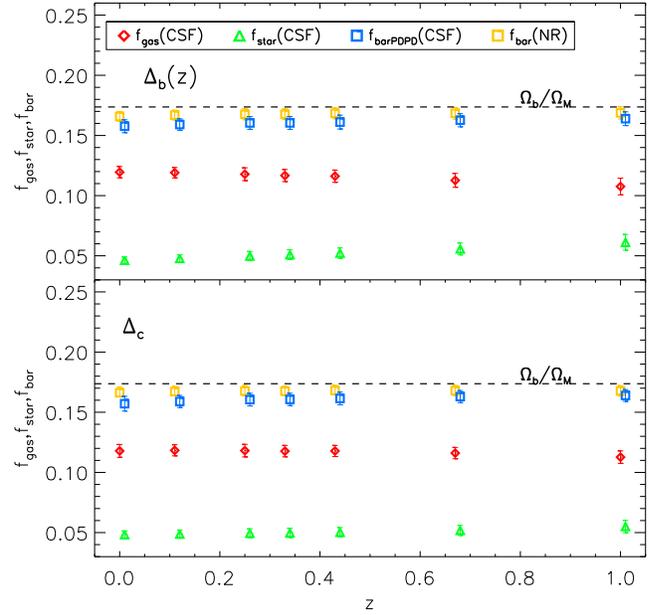}
\caption{Evolution of $f_{gas}$ (red diamonds), $f_{star}$ (green triangles) and $f_{bar}$ (yellow squares) of CSF clusters and $f_{bar}$ of NR clusters (blue squares) for $\Delta _b$=1500($z$) (top panel) and $\Delta _c$=500 (bottom panel).}
\label{500}
\end{figure}

The evolution of the same baryonic component fractions is shown in
Fig. \ref{500} but  for a larger integration domain defined by an overdensity: $\Delta_b = 1500(z)$ and $\Delta_c = 500$ respectively, corresponding to $\sim R_v/2$. In this case there are neither  any  relevant differences between the two overdensities used to define the radii, nor there is  a significant evolution with redshift  observed.   Therefore, we can confidently  assume   the values at  $z=0$ for  $\Delta_c$=500   to be representative (within 
10\% ) of the gas fraction values for clusters  up to $z=1$: 
$f_{gas}$=0.118$\pm$0.005, $f_{star}$=0.048$\pm$0.003,
$f_{bar}$=0.166$\pm$0.004. For NR clusters we have  $f_{bar}$=0.157$\pm$0.006.

\begin{figure}
\centering\includegraphics[angle=0,width=9cm]{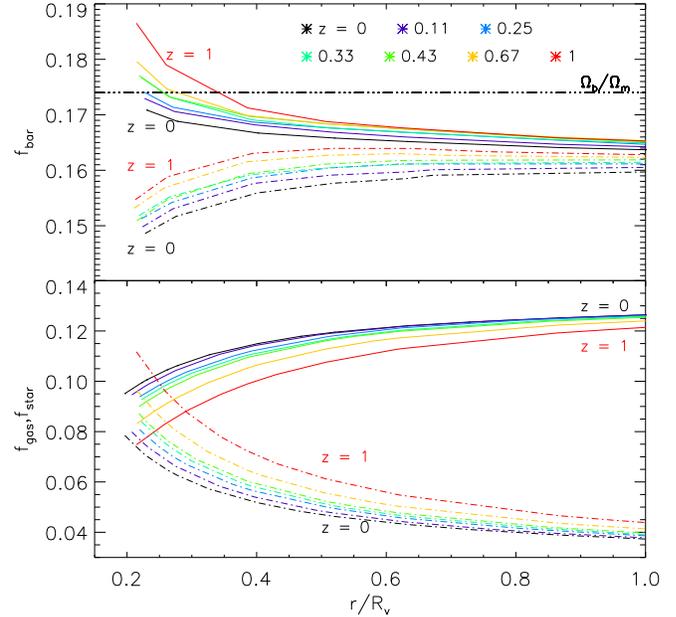}
\caption{Top panel: Average radial profiles,  in units of the  virial radius, of  $f_{bar}$ for CSF clusters (continuous lines) and for NR cluster (dashed) at different redshifts from $z$=0 (black) to $z$=1(red). Bottom panel: $f_{gas}$ (continuous lines) and $f_{star}$ (dashed lines) profiles for CSF  clusters  only,  from $z$ = 0 (black) to $z$ = 1 (red).}
\label{prof}
\end{figure}

\begin{figure}
\centering\includegraphics[angle=0,width=9cm]{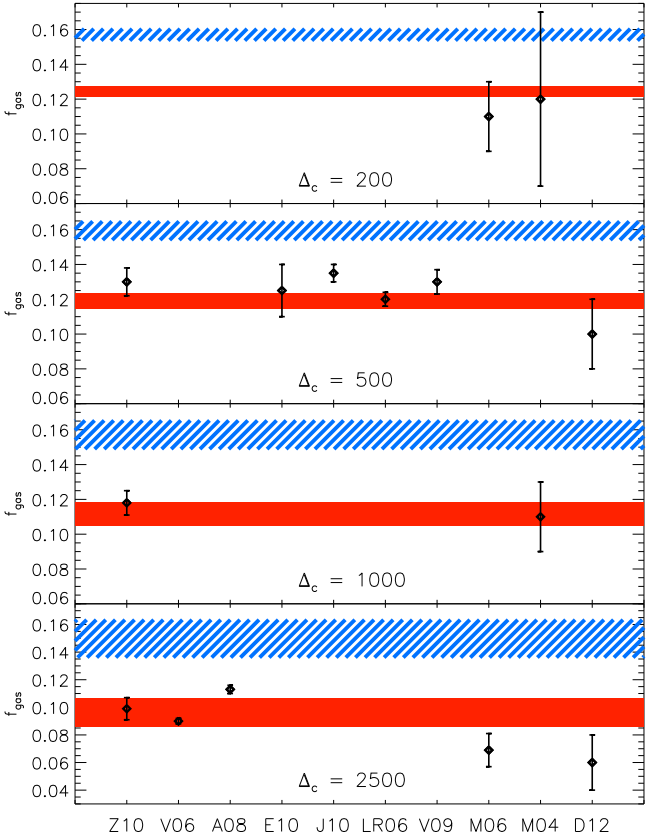}
\caption{Comparison among gas fractions of  MUSIC-2 clusters at
  different overdensities and observational estimates. 
Blue dashed area shows the   $1 \sigma$  dispersion of
  $f_{gas}$  derived from the volume limited  NR   MUSIC-2  clusters
  while the red solid area corresponds to the  CSF MUSIC-2 clusters. The
$x$-axis  denotes the different observations:  Z10
\citep{ZHANG10}, V06 \citep{VI06}, A08  \citep{AL08}, E10 \citep{ET10},
J10  \citep{JU10}, LR06,  \citep{LAROQUE06}, V09 \citep{VI09}, D12
\citep{DJF12}, M04 \citep{MGAN04},  M06 \citep{MGAN06}.}
\label{locuss}
\end{figure}

Fig. \ref{prof} shows the average radial (normalized to the virial radius) cluster profiles   of gas, star and baryonic fractions for all the redshift considered in this analysis.  
 An evolution of the baryon properties with redshift is clearly seen in this plot, which is  more relevant at smaller $R/R_v$. The gas fraction
increases from inner to  outer regions and shows higher values at
low redshifts.  On the contrary, the star fraction decreases when moving
towards the virial radius and is bigger at higher redshifts, showing that star formation is preferentially located at the central parts of clusters at low redshift.  The baryon
fraction decreases when reaching the cluster outskirts for CSF clusters
(with mean values higher than the cosmic ratio) while the NR clusters
show the opposite behaviour. In both physical configurations the mean
baryon fraction becomes smaller when moving from $z$ = 1 to $z$ = 0.  The evolution of $f_{bar}$  at large fractions of the virial radius is very mild for CSF  clusters  and converges towards  a nearly constant gas fraction  for  $\Delta_c = 500$.

As a summary of the previous discussion we conclude that  using $\Delta_b$($z$)  to define the aperture radius   at different redshift allows to reliably compare the same region ( with respect to the virial radius of the cluster at each redshift) and thus is the  best choice  when  studying  redshift evolutions. On the other hand, by selecting the  radius  based on fixed  $\Delta_c$  values allows us to  use a  $f_{gas}$ which is  nearly constant with redshift.

We  now compare   the values of the  gas fractions  derived from our simulations   with those measured in  observations.

  In Fig.\ref{locuss} we  show a compilation of published   $f_{gas}$ estimates from  observations at different overdensities and our numerical results for the corresponding overdensities, both from the NR and CSF MUSIC clusters within the 1$ \sigma$   dispersion  values. 

 Our results are consistent  with the gas fractions  that have been measured at  3 overdensities ($\Delta_c= 2500, 1000, 500$)  by the LoCuSS project \citep{ZHANG10}.
At $\Delta_c= 500$ we also found consistency with   $f_{gas}$ estimated by  \cite{ET10}, \cite{JU10},
\cite{LAROQUE06}, \cite{VI09}, \cite{MGAN06}, \cite{MGAN04} and
\cite{DJF12}.

At $\Delta_c$=2500 our results are still consistent with \cite{VI06} and \cite{MGAN06}, while \cite{AL08} shows a higher value of the gas fraction. 

At $\Delta_c$=200  our clusters are also compatible with   the two only measurements at these large radius, \cite{MGAN06}, \cite{MGAN04}, but due to the larger error bars  they are less restrictive. 

\textbf{The star fraction in MUSIC-2 CSF clusters is on average overestimated in comparison with observations. While a value lower than 1 per cent at $\Delta_c$ = 200 is reported in  \cite{GONZALEZ07}, \cite{BODE09}, \cite{GIODINI09} and \cite{ANDREON10}, we derive a star fraction at $\Delta_c \leq$ 500 of the order of 4-5 per cent.}

\section{SZ cluster scaling relations} \label{sec:theory}
The correlation of the SZ signal with other cluster properties was initially studied by considering the central comptonization parameter, $y_0$,
(see \citealt{COORAY99} and \citealt{MC03}) but the inadequacy of a simple $\beta$-model for the ICM 
spatial distribution was in short clear. 
For the first time \cite{BENS04} applied the Compton $y$-parameter integrated within a redshift-independent density contrast, $\Delta_c$=2500, in a SZ scaling relation for 15 clusters observed with SuZIE II. As the $y_0$ parameter gives only local information about the status of the cluster, the integrated $Y$ constitutes a global property, strongly bound to the energy of the cluster \citep{MRK09} and less model-dependent.

The comparison between quantities integrated within a constant critical overdensity is now widely employed in 
X-ray and SZ scaling relations analysis at overdensities $\Delta_c$=2500 (\citealt{COMIS11}; \citealt{BONA2008}) and $\Delta_c$ = 500 (\citealt{ARNAUD10}; \citealt{PLANCKb}, \citealt{AND11}). 
The great benefit, distinctive in SZ observations of ignoring cluster redshift to correct the redshift 
dependence of the overdensity, is evident. 

In order to give an estimate of the integrated $Y$ of MUSIC clusters, we have produced simulated maps of the thermal SZ effect, integrating along the projected radius to extract the value of $Y$ at fixed redshift and overdensity.

\subsection{Cluster $y$-map generation} \label{sec:theory}
The maps of the Compton $y$-parameter  are generated with the implemented procedure employed in \cite {INES09}. The simulation provides for each $i$-th SPH particle: position (\textbf{r$_i$}), velocity (\textbf{v$_i$}), density (\emph{$\rho _i$}), internal energy (\emph{U$_i$}) and SPH smoothing length (\emph{h$_i$}). In CSF simulations we also find the number of ionized electrons per hydrogen particle (\emph{N$_{e,i}$}) and metallicity \emph{Z$_i$}: using this additional information we could achieve a more detailed estimate of electron temperature \emph{T$_e$} (per gas particle) as:
\beq
T_e=(\gamma -1)U m_p \frac{\mu}{k_B}
\eeq
where $m_p$ is the proton mass, $\gamma$ the polytropic coefficient (set to 5/3 for monatomic gases) and $\mu$ the mean molecular weight of electrons, that we estimate as
\beq
\mu = \frac{1-4Y_{He}}{1+\gamma+N_e}
\eeq
($Y_{He}$ is the nuclear helium concentration).
The number electron density (in cm$^{-3}$) is derived as
\beq
n_e = N_e \rho_{gas}\bigg(\frac{1-Z-Y_{He}}{m_p}\bigg)
\eeq
(The density $\rho_{gas}$ of the gas particles of CSF particles has been
corrected in order to take into account the multiphase model adopted in
the simulation)

We are interested only in the thermal component of the SZ effect. This is quantified in the Compton $y$-parameter:
\beq
y\quad = \quad \int n_e \frac{k_BT_e}{m_ec^2}\sigma_T dl
\eeq
The integral can be discretized along the line of sight (\emph{los}) as

\beq
y \simeq \frac{k_B\sigma_T}{m_ec^2}\sum_in_{e,i}T_{e,i}W_p(\mid {\bf r_i}-{\bf r_{com}}\mid,h_i),
\eeq
where the tSZ signal associated to each particle is spread to the surrounding area  by the mass profile m$_{gas}$W$_p$($\mid${\bf r$_i$ - r$_{com}$}$\mid$, $h_i$), where $W_p$ is the projection of  the normalized spherically symmetric spline kernel $W$ used in the simulations:

\beq
W(x,h_i)=\frac{8}{\pi h_i^3} \left\{ \begin{array}{cc}
 1-6x^2+6x^3, &\quad 0 \leq x \leq 0.5\\
 2(1-x)^3 ,  & \quad 0.5 \leq x \leq1\\
 0,  & \quad x>1
       \end{array} \right.
\eeq

\subsection{SZE scaling relations} \label{sec:theory}
The integrated Compton $y$-parameter, $Y$, appears to be a very good
cluster mass proxy. With the aim to properly employ SZ observations as
mass finder, it is fundamental to accurately calibrate the  $Y-M$ scaling
relation.

As the integrated Compton parameter Y is a very good proxy of the clusters mass, it is fundamental to study the Y-M scaling relation.
The $Y$ parameter is estimated as follows:
\begin{equation}
Y \equiv \int _{\Omega} yd\Omega=D_A^{-1} \bigg(\frac{k_b\sigma_T}{m_ec^2}\bigg)\int_0^\infty dl \int _An_eT_edA
\end{equation}
where $D_A$ is the angular distance and the integration is performed
inside the solid angle, $\Omega$, equal to an area $A$ in the projected map,
corresponding to an overdensity $\Delta$.

\begin{figure}
\centering\includegraphics[angle=0,width=9cm]{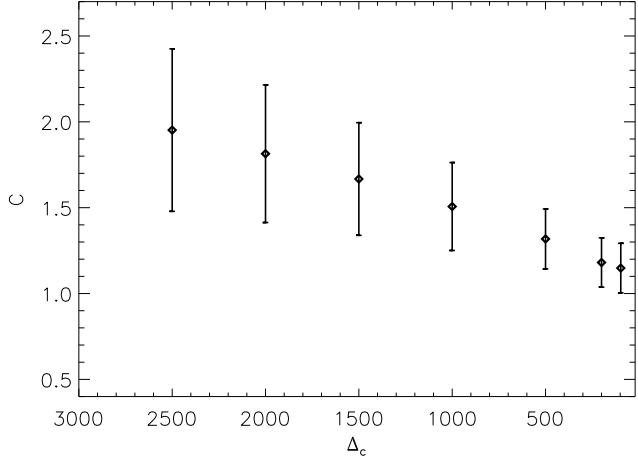}
\caption{Behavior of the $C$=$Y_{C}D_A^2$/$Y_S$ parameter with critical overdensity: each value corresponds to the mean value calculated over all the redshifts analyzed. The error bars represent the mean error calculated from the mean values of $C$ over all redshifts.}
\label{C}
\end{figure}

Adopting a self-similar model to describe clusters of galaxies (assuming hydrostatic equilibrium and isothermal distribution for dark matter and gas particles) the gas temperature is related to the mass by a simple power law:
\begin{equation}
T_e \propto (M_{\Delta}E(z))^{2/3}\label{t_e}
\end{equation}
where, as we said in the previous section, the total mass is calculated inside  the overdensity radius $r_{\Delta}$ we are considering and $E(z)$ the Hubble constant normalized to its present day value defined as $E(z)$=($\Omega _{M}(1+z)^3+\Omega _{\Lambda}+\Omega_k(1+z)^2)^{1/2}$ .
Due to the hypothesis of isothermal equilibrium, $Y$ is proportional to the integral of the electron density {\itshape n$_e$} over a cylindrical volume, leading to:
\begin{equation}
Y_CD_A^2\propto T_e\int n_edV\propto M_{gas}T_e=f_{gas}M_{tot}T_e
\end{equation}
Considering Eq.(\ref{t_e}), we find a relation  that depends directly  on  $M_{tot}$ and $f_{gas}$:
\begin{equation}
Y_CD_A^2\propto f_{gas}M_{tot}^{5/3}E(z)^{2/3}\label{YD}
\end{equation}
To fit a spherical quantity, like the total energy of a cluster, it is better to introduce the spherical integrated $Y_S$ parameter.
This is defined as the Compton-y parameter integrated inside a spherical volume with a radius equal to $r_\Delta$. The Eq.(\ref{YD}) turns out to be:
\begin{equation}
Y_{S,\Delta_c}=\frac{\sigma _T}{m_ec^2}\frac{\mu}{\mu _e}\bigg(\frac{\sqrt{\Delta_c}GH_0}{4}\bigg)^{2/3}E(z)^{2/3}f_{gas,\Delta}M_{tot,\Delta}^{5/3}\label{YS}
\end{equation}
where with $Y_{S,\Delta}$ we denote  the $Y$ parameter integrated over a cluster centred  sphere of radius $r_{\Delta}$. Here, we are assuming a constant {\itshape f$_{gas}$} independent on  the cluster  mass  for each overdensity.
We calculate from the map the $Y_{\Delta}$ parameter at the considered overdensity integrating along all the pixels of the projected map falling inside the projected radius $r_{\Delta}$:
\begin{equation}
Y_{S,\Delta} =\sum_{i,r<r_{\Delta}}y_i \times d\l^2_{pix}
\end{equation}
In order to take into account the different integration domain, we define the $C$ parameter as:
\beq
C = \frac{Y_CD_A^2}{Y_S}
\eeq
In Fig \ref{C} we give an estimate of $C$ from the values of $Y_S$ and $Y_CD_A^2$ computed from our clusters at different redshifts and overdensities. We find  that $C$ is constant
with redshift and grows with overdensity. Available values in literature are 
consistent with ours: at $\Delta_c$=2500 we found
$C$=2.0$\pm$0.5, in agreement with the estimate of $C\simeq$ 2 given by
\cite{BONA2008} and $C$=1.67$\pm$0.80 given by \cite{COMIS11}, while at
$\Delta_c$=500 we have $C$=1.3$\pm$0.2, compatible with \cite{AND11}.  From now on
we will  denote the spherical $Y_S$ parameter simply as  $Y$.

The $Y-M$ scaling relation at a fixed overdensity is studied performing a  best fit of
\begin{equation}
Y_{\Delta} = 10^B\bigg(\frac{M_{\Delta}}{\hMsun}\bigg)^AE(z)^{2/3}[h^{-2}Mpc^{2}]\label{eqYM}
\end{equation}
where $M_{\Delta}$ is the total mass calculated inside the sphere of radius $r_{\Delta}$ that we are considering and $B$ the normalization, defined as:
\begin{equation}
B = \log \frac{\sigma _T}{m_ec^2}\frac{\mu}{\mu _e}\bigg(\frac{\sqrt{\Delta_c}GH_0}{4}\bigg)^{2/3}+ \log f_{gas},  \label{B}
\end{equation}
contains all the constant terms of the first part of the right hand of Eq.(\ref{YS}).  We remind that, if the clusters follow the self-similar model,  then the slope $A$ is expected to be $5/3$.

We are also interested in the inverse scaling relation $M-Y$,  which is more easily  applicable  to observational  clusters (the mass should be inferred from $Y$ measurements):

\begin{equation}
M_{\Delta} = 10^{B*}\bigg(\frac{Y_{\Delta}}{h^{-2}Mpc^{2}}\bigg)^{A*}E(z)^{-2/5}[\hMsun]
\end{equation}

We study these scaling relations in the log-log  space, in the form $\log \Upsilon = B + A\log X$, where in the case of the $Y-M$ relation we have $X=M$ and $\Upsilon=YE(z)^{\kappa}f_{gas}^{-1}$: $\kappa$ is fixed to its expected self-similar scaling with $z$ ($\kappa=-2/3$) and in this case $f_{gas}$ is not included in the normalization parameter $B$.

\begin{figure}
\centering\includegraphics[angle=0,width=8cm]{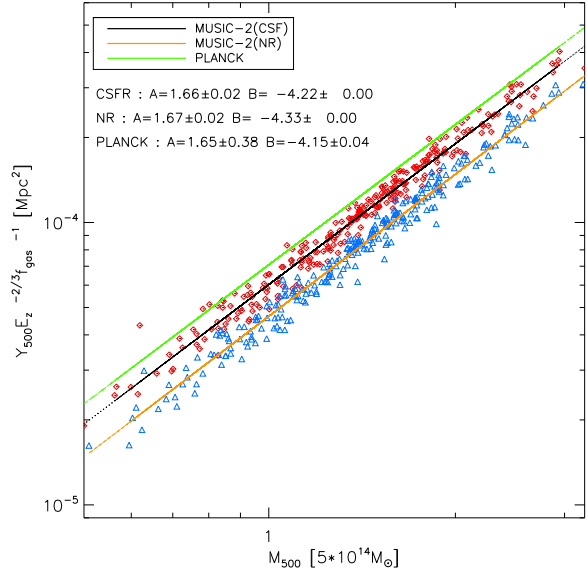}
\caption{$Y-M$ scaling relation at  $z$ = 0,  $\Delta_c$ = 500. CSF clusters are shown as red diamonds (the best fit as a black solid  line) and NR as blue triangles (best fit as an orange solid line). The green line shows the best fit \textbf{from recent Planck data employing cluster masses as derived by weak lensing observations, \citep{PLANCK12}.}}
\label{YM}
\end{figure}

Figure \ref{YM} shows the $Y-M$ scaling relation of our dataset at $\Delta_c = 500$ for both physical configurations. We choose to consider a value of $f_{gas}$ constant at fixed overdensity and redshift, assigning to all clusters the mean value of $f_{gas}$ as:
\begin{equation}
f_{gas}(\Delta,z)=\sum^N_{i=1}\frac{f_{gas,i}(\Delta,z)}{N}\label{frac}
\end{equation}
where $N$ is the number of clusters in the considered subset.

CSF clusters show higher values than NR clusters, as the relation
depends on the inverse of $f_{gas}$, which is bigger for NR clusters, (see Section \ref{sec:bar}).

\begin{figure}
\centering\includegraphics[angle=0,width=9cm]{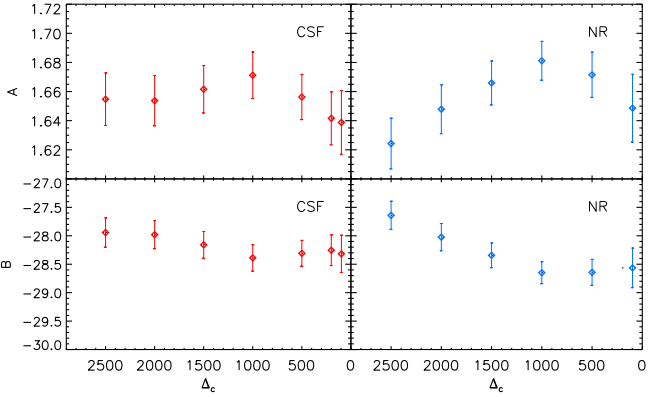}
\caption{  
The best-fit slope $A$ (top panels) and normalization $B$ (bottom panels) of the $Y-M$ scaling relation as a function of overdensity  for  $z=0$  CSF  (left panel) and NR (right panel) MUSIC-2 clusters.}
\label{abYM}
\end{figure}

The values of $A $ and $B$   at $z$ = 0 with  their  1-$\sigma$ errors  are
listed in Table \ref{tabYM}  and shown in Fig.\ref{abYM}.  At all
the overdensities considered,  the values of the slope $A $  seem to agree, within  their  errors,   with the self-similar prediction of $5/3$. NR clusters
seem to show a lower slope at high overdensities and CSF clusters an
higher one, matching both with the same self-similar value at
$\Delta_c$=500. The value of $B$ seems to stay  almost constant at all
overdensities.

Leaving aside  a possible variation with redshift, which is studied  in section \ref{sec:results}, we can therefore express the $Y-M$  scaling relation at $\Delta_c$ = 500 as:
\begin{equation}
Y_{500} = 10^{-28.3\pm 0.2}\bigg(\frac{M_{500}}{\hMsun}\bigg)^{1.66\pm 0.02}E(z)^{2/3}[h^{-2}Mpc^{2}]
\end{equation}
We estimate the scatter on the $Y-M$ relation 
 by computing  the sum of the residuals as follows:
\begin{equation}
\sigma_{\log_{10}Y} = \sqrt{\frac{\sum^N_{i=1}[\log Y_{\Delta,i}-(A \log M_{\Delta,i}+B)]^2}{N-2}}\label{sigma}
\end{equation}
As shown in Table  \ref{tabYM}, the scatter on the $Y$ estimate is
about 5 per cent at all overdensities.

\begin{figure}
\centering\includegraphics[angle=0,width=8cm]{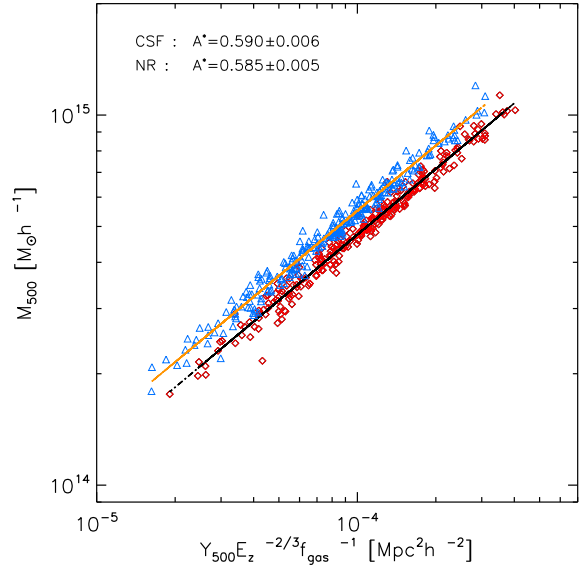}
\caption{$M-Y$ scaling relation at $z = 0$   for $\Delta_c = 500$: CSF clusters are shown as red diamonds (the best fit as a black line) and NR as blue triangles (best fit as an orange line).}
\label{MY}
\end{figure}

\begin{figure}
\centering\includegraphics[angle=0,width=9cm]{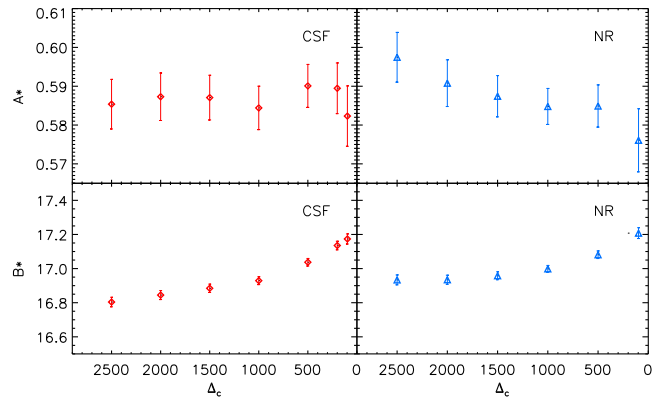}
\caption{Same as  Fig  \ref{abYM} but for the  $M-Y$ scaling relation.}
\label{abMY}
\end{figure}

In the same way we also calculate the $M-Y$ scaling relation.  Fig. \ref{MY} shows the relation at $z$ = 0 for  $\Delta_c$=500, for which the self-similar model predicts a slope $A^*$ = 3/5. Table  \ref{tabYM} also lists the best fit  values of  $A^*$  at $z$ = 0, which seem to be  rather independent on  overdensity and to agree within the errors  with the self-similar prediction.  In Fig.\ref{abMY} we show the evolution of $A^*$ and $B^*$ with overdensity. We can then express the best-fit $M-Y$ scaling relation at $\Delta_c$ = 500 as:
\begin{equation}
M_{500} = 10^{17.0\pm 0.1}\bigg(\frac{Y_{500}}{h^{-2}Mpc^2}\bigg)^{0.59\pm 0.01}E(z)^{-2/5}[\hMsun]
\end{equation}

\begin{figure*}
\centering\includegraphics[angle=0,width=15cm]{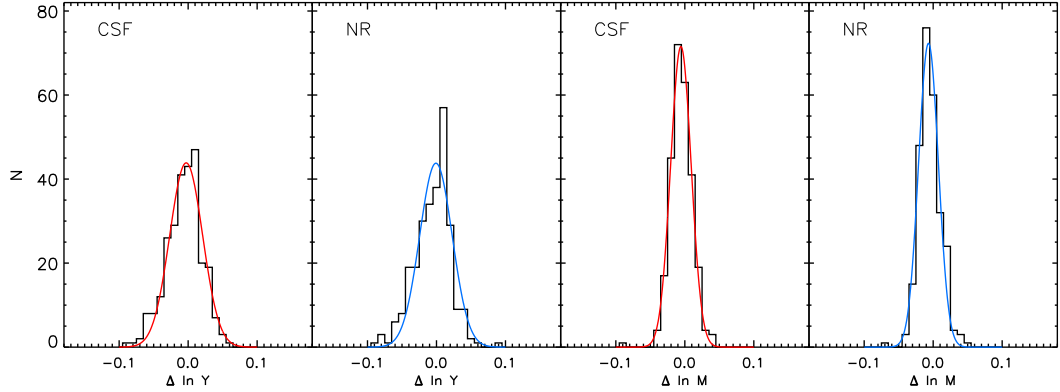}
\caption{Distribution of the  residuals  of the $Y-M$ (in $\log Y$, left panels) and $M-Y$ (in $\log M$, right panels) scaling relations at $z=0$ for $\Delta_c$ = 500. The red (CSF) and blue (NR) lines represent the best fit Gaussian curves to the  residual distribution.}
\label{resYM}
\end{figure*}

Moreover, the $M-Y$ relation seems to show a smaller scatter (3 per cent)  than
the $Y-M$. The difference between the two relations is shown in Fig.$\ref{resYM}$, in which we show the distribution of the residuals at $\Delta_c$ = 500. Both distributions are well fitted by  a Gaussian with zero mean and different widths. The distributions of residuals for the $M-Y$   is narrower (i.e. smaller $\sigma$ ) than the corresponding one  for $Y-M$. 

Therefore, we can  establish that the MUSIC-2 clusters show $Y-M$ and
$M-Y$ scaling relations  that are well consistent with the self-similar model
both at high and low overdensities. This result confirms, if necessary,
the validity of the self-similar model in using the integrated $Y$
parameter as a proxy of the mass and,  at the same time,  demonstrates the
reliability of the MUSIC cluster dataset for the  study  of scaling relations. No
significant discrepancies can be pointed  out between the two relations,
$Y-M$ and $M-Y$.  Therefore we  will focus  all further  analyzes   described in the
next sections only on the first one.

\subsection{Comparison with observations} 
Several scaling-laws  have been derived from different  SZ observations 
(\citealt{BONA2008}; \citealt{SPT2011A}; \citealt{ARNAUD10}; \citealt{COMIS11}; \citealt{PLANCKc}; \citealt{AND11}).  Here,  our numerical results are compared with the 
 best fit scaling relation $Y_{500} - M_{500}$ based \textbf{on 19 galaxy clusters detected at high signal-to-noise in the Planck all-sky data set, whose masses have been derived using weak lensing \citep{PLANCK12}: MUSIC-2 clusters well fit the Planck scaling relation, with a better agreement shown by the CSF dataset, as can be seen in Fig.\ref{YM}.}

\begin{figure}
\centering\includegraphics[angle=0,width=8cm]{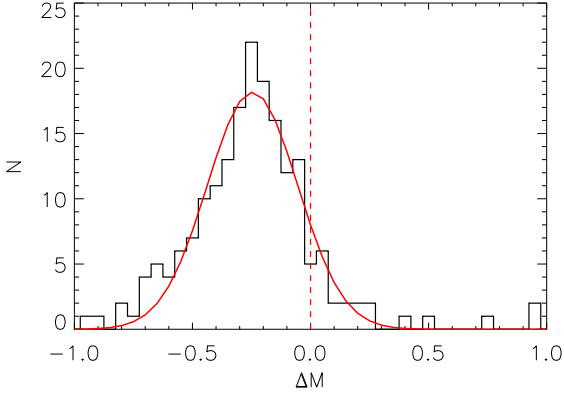}
\caption{Distribution of  the hydrostatic mass  bias of  $M_{500}$ for  CSF clusters at $z$ = 0. The red curve represents the Gaussian curve which best fits the distribution. The Gaussian is centered at $\Delta M = -0.25$  with a  width of $\sigma=0.43$.}
\label{HSE}
\end{figure}

The masses assigned to the  MUSIC clusters   are of course free of systematic errors since they  are the true masses computed  from the simulations.  We have  check  how the numerical SZ scaling relation would be affected  if we assume the hypothesis of hydrostatic equilibrium  (HSE), to derive the mass of the clusters, as is usually done in observations. The HSE assumption constitutes one of the most significant uncertainties in the derivation of the  observational scaling relations.  We calculate  the HSE cluster mass  at a given overdensity  as:
\begin{equation}
M_{HSE,\Delta}(< r) = -\frac{kTr}{G\mu m_H}\bigg( \frac{d\ln \rho}{d\ln r} + \frac{d\ln T}{d\ln r}\bigg)
\end{equation}
We used the mass weighted temperature profile  of the gas   using Eq.(\ref{mw}) to calculate the gas particle temperature.   Although the spectroscopic like-temperature would have  been  more fitted to compare with X-ray observations, we do not expect that the conclusions  could be significantly  different.  The distribution of the  Hydrostatic Mass Bias (HMB), quantified  as $\Delta M$=($M_{500,HSE}$-$M_{500}$)/$M_{500}$, can be seen in Fig. \ref{HSE}.  A  systematic underestimation of  25 per cent  of the  true  mass by  using the hydrostatic mass is found, with a dispersion of $\sigma = 0.43$. The result is consistent with  recent numerical simulations that also find that the   HSE assumption underestimates the true mass up to 20  per cent (\citealt{RASIA06}; \citealt{KAY07};  \citealt{NKV07}; \citealt{NVK07}; \citealt{PIFFA08}; \citealt{AMEGLIO09}; \citealt{LAU09}). \textbf{On the other hand, the estimation of HSE masses derived in \cite{PLANCKb} shows an opposite behavior: in this case they have found that the HSE hypothesis introduces an overestimation of about 20 per cent with respect to the mass calculated via weak lensing \citep{PLANCK12}. They have attributed this discrepancy to the errors  in the  mass concentration
measurements and to an offset in the cluster centers between X-rays and
lensing observations.} A more detailed analysis of the HMB, 
especially at high overdensity regions  where a significant contribution 
from non-thermal pressure is present even  in relaxed clusters, is under way and it will be the topic of a  further  paper.

\begin{table*}
\begin{center}
\begin{tabular}{|c|c|c|c|c|c|c|c|c|}
\hline
 & \textbf{$\Delta_c$} & \textbf{2500} & \textbf{2000} & \textbf{1500} & \textbf{1000} & \textbf{500} & \textbf{200} & \textbf{98$v$} \\
\hline
\textbf{CSF} & & & & & & & & \\
\hline
& \textbf{A} & 1.65 $\pm$ 0.02 & 1.65 $\pm$ 0.02 & 1.66 $\pm$ 0.02 & 1.67 $\pm$ 0.02 & 1.66 $\pm$ 0.02 & 1.64 $\pm$ 0.02 & 1.64 $\pm$ 0.02 \\
$Y-M$ & \textbf{B} & -27.9 $\pm$ 0.3 & -28.0 $\pm$ 0.2 & -28.2 $\pm$ 0.2 & -28.4 $\pm$ 0.2 & -28.3 $\pm$ 0.2 & -28.3 $\pm$ 0.2 & -28.3 $\pm$ 0.3 \\
& \textbf{$\sigma_{\log Y}$} & 0.05 & 0.05 & 0.05 & 0.04 & 0.04 & 0.04 & 0.05 \\
\hline
& \textbf{A*} & 0.585$\pm$ 0.006 & 0.587$\pm$ 0.006 & 0.587$\pm$ 0.006 & 0.584$\pm$ 0.006 & 0.590$\pm$ 0.005 & 0.589$\pm$ 0.007 & 0.582$\pm$ 0.008 \\
$M-Y$ & \textbf{B*} & 16.8 $\pm$ 0.03 & 16.8 $\pm$ 0.02 & 16.9 $\pm$ 0.02 & 16.9 $\pm$ 0.02 & 17.0 $\pm$ 0.02 & 17.1 $\pm$ 0.02 & 17.2 $\pm$ 0.03 \\
& \textbf{$\sigma_{\log M}$} & 0.03 & 0.03 & 0.03 & 0.03 & 0.02 & 0.03 & 0.03 \\
\hline
\textbf{NR}& & & & & & & & \\
\hline
& \textbf{A} & 1.62 $\pm$ 0.02 & 1.65 $\pm$ 0.02 & 1.67 $\pm$ 0.02 & 1.68 $\pm$ 0.02 & 1.67 $\pm$ 0.02 & 1.66 $\pm$ 0.02 & 1.65 $\pm$ 0.02 \\
$Y-M$ & \textbf{B} & -27.6 $\pm$ 0.2 & -28.0 $\pm$ 0.2 & -28.3 $\pm$ 0.2 & -28.6 $\pm$ 0.2 & -28.6 $\pm$ 0.2 & -28.3 $\pm$ 0.2 & -28.5 $\pm$ 0.3 \\
& \textbf{$\sigma_{\log Y}$} & 0.06 & 0.05 & 0.04 & 0.04 & 0.04 & 0.04 & 0.05 \\
\hline
& \textbf{A*} & 0.597$\pm$ 0.006 & 0.591$\pm$ 0.006 & 0.587$\pm$ 0.005 & 0.585$\pm$ 0.005 & 0.585$\pm$ 0.005 & 0.598$\pm$ 0.006 & 0.576$\pm$ 0.008 \\
$M-Y$ & \textbf{B*} & 16.9 $\pm$ 0.03 & 16.9 $\pm$ 0.02 & 17.0 $\pm$ 0.02 & 17.0 $\pm$ 0.02 & 17.1 $\pm$ 0.02 & 17.2 $\pm$ 0.02 & 17.2 $\pm$ 0.03 \\
& \textbf{$\sigma_{\log M}$} & 0.03 & 0.03 & 0.03 & 0.02 & 0.02 & 0.02 & 0.03 \\
\hline
\end{tabular}
\end{center}
\caption{Best fit slope, normalization and scatter of $Y-M$ (A, B, $\sigma_{\log Y}$) and $M-Y$ (A*, B*, $\sigma_{\log M}$) scaling relations for CSF and NR clusters.}
\label{tabYM}
\end{table*}

\begin{figure}
\centering\includegraphics[angle=0,width=7cm]{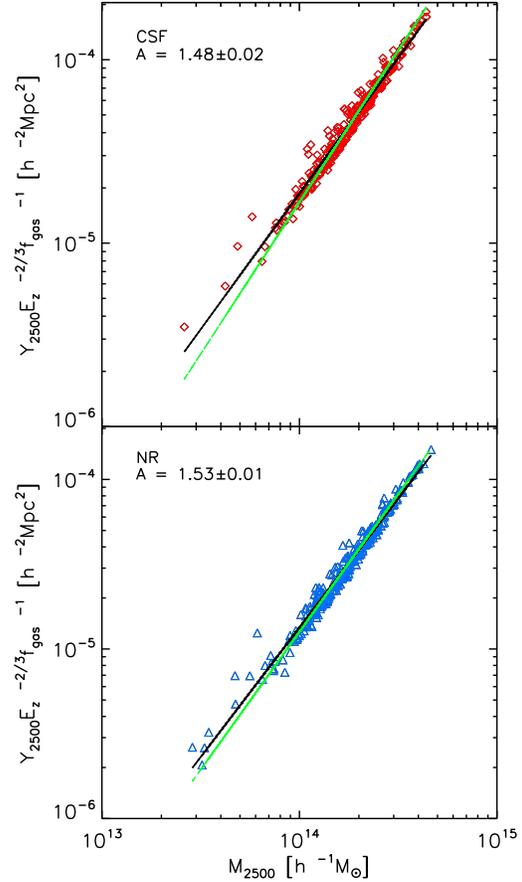}
\caption{$Yf_{gas}-M$ scaling relation at $z$ = 0, $\Delta_c$ = 2500 for CSF clusters (top panel) and NR clusters (bottom panel). The black line shows the best-fit relation, the green line the best-fit relation of the $Y-M$ scaling relation (with constant $f_{gas}$).}
\label{Yfgas}
\end{figure}

\section{The effect of the gas fraction on the Y-M scaling relations}
In  the previous section we  assumed a constant  gas fraction for all clusters used to  fit the 
$Y-M$ scaling law at a given redshift and overdensity. This assumption is a common observational approach in  scaling laws study due to the large uncertainty occurring in single cluster
measurements. In numerical simulations  we can easily estimate the gas fraction of
each single object and  investigate in more detail its impact on the $Y-M$ scaling relation.

In  this section we leave the value of $f_{gas}$ free. Therefore,  using  Eq.(\ref{eqYM})  we can derive   the modified scaling relation:
\begin{equation}
Y_{\Delta}f_{gas}^{-1} = 10^{B1}\bigg(\frac{M_{\Delta}}{\hMsun}\bigg)^{A1}E(z)^{2/3}[h^{-2}Mpc^{2}]\label{eqYf}
\end{equation}

\begin{figure*}
\centering\includegraphics[angle=0,width=18cm]{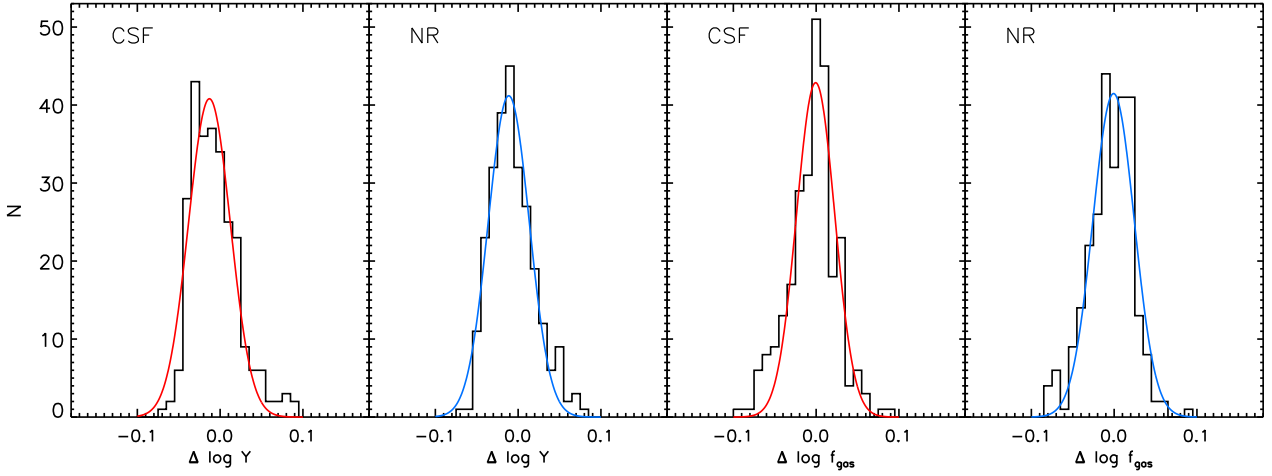}
\caption{left panel:  the distribution of residuals for $Yf_{gas}-M$ at  $z=0$ and    $\Delta_c$=2500.  Right panel: same as left panel but for the  $f_{gas}-M$ scaling relation. The red (CSF) and blue (NR) lines represent
the Gaussian curves which best  fit  the residual distribution.}
\label{YMF}
\end{figure*}

In Table  \ref{tabMF} we present the  best fit values  of  the  $Yf_{gas}^{-1}-M$ relation at $z$ = 0,  for overdensities  in the range  $\Delta_c=2500$  to $\Delta_{v,c}$.     We note that   the slope  is steeper  and the normalization is  larger (in absolute value) than  the $Y-M$  with fixed  $f_{gas}$, as can be seen in Fig. \ref{Yfgas}, and the distribution of residuals is wider and less regular (Fig.\ref{YMF}).   Since we saw  before that $Y-M$   agrees very well with the  self-similar  predictions,  it follows that the departure from self similarity can be attributed to the dependence of $f_{gas}$ on   the cluster total  mass.  This effect seems to be independent on redshift but very sensitive on  the overdensity value:  core regions present   a larger deviation from the self-similar values   than  halo outskirts.

\begin{table*}
\begin{center}
\begin{tabular}{|c|c|c|c|c|c|c|c|c|}
\hline
 & \textbf{$\Delta_c$} & \textbf{2500} & \textbf{2000} & \textbf{1500} & \textbf{1000} & \textbf{500} & \textbf{200} & \textbf{98$v$} \\
\hline
\textbf{CSF} & & & & & & & & \\
\hline
\hline
& \textbf{A1} & 1.48 $\pm$ 0.02 & 1.51 $\pm$ 0.01 & 1.55 $\pm$ 0.01  & 1.59 $\pm$ 0.01  & 1.61 $\pm$ 0.01 & 1.62 $\pm$ 0.01  &  1.63 $\pm$ 0.02 \\
$Y-M_{gas}$ & \textbf{B1} & -25.5$\pm$ 0.2 & -25.9 $\pm$ 0.2  & -26.5 $\pm$ 0.2  & -27.2 $\pm$ 0.2 & -27.7 $\pm$ 0.2  & -28.0 $\pm$ 0.2 &  -28.2 $\pm$ 0.3  \\
\hline
& \textbf{A2} & 0.17$\pm$ 0.01 & 0.15 $\pm$ 0.01  & 0.11$\pm$ 0.01  & 0.08 $\pm$ 0.01 & 0.04 $\pm$ 0.01  & 0.02 $\pm$ 0.01  &  0.01 $\pm$ 0.01\\
$f_{gas}-M$ & \textbf{B2} & -3.5 $\pm$ 0.2 & -3.1 $\pm$ 0.2 & -2.6 $\pm$ 0.2 & -2.2 $\pm$ 0.1 & -1.6 $\pm $0.1 & -1.2 $\pm $0.1 &  -1.0 $\pm$ 0.1\\
\hline
\textbf{NR}& & & & & & & & \\
\hline
& \textbf{A1} & 1.52 $\pm$ 0.01 & 1.56 $\pm$ 0.01 & 1.59 $\pm$ 0.01 & 1.63 $\pm$ 0.01  & 1.66 $\pm$ 0.01  & 1.66 $\pm$ 0.01  & 1.65 $\pm$ 0.01  \\
$Y-M_{gas}$ & \textbf{B1} & -26.2 $\pm$ 0.02 & -26.8 $\pm$ 0.02 & -27.3 $\pm$ 0.02 & -27.9 $\pm$ 0.01  & -28.4 $\pm$ 0.01  & -28.5 $\pm$ 0.01  & -28.6 $\pm$ 0.01 \\
\hline
& \textbf{A2} & 0.10 $\pm$ 0.01 & 0.09 $\pm$ 0.01 & 0.07 $\pm$ 0.01 & 0.05 $\pm$ 0.01 & 0.02 $\pm$ 0.01  & -0.03 $\pm$ 0.01  &  -0.04 $\pm$ 0.01\\
$f_{gas}-M$ & \textbf{B2} & -2.2 $\pm$ 0.2 & -2.1 $\pm$ 0.2 & -1.9 $\pm$ 0.2 & -1.6 $\pm$ 0.1 & -1.0 $\pm$ 0.1 & -0.8 $\pm$ 0.1 & -0.7 $\pm$ 0.1 \\
\hline
\end{tabular}
\end{center}
\caption{Best fit slope, normalization and scatter of $Yf_{gas}-M$ (A1,B1) and $f_{gas}-M$ (A2, B2) scaling relations for CSF and NR clusters.}
\label{tabMF}
\end{table*}

  We now study  the  relation between $f_{gas}-M$  in the  form:
\begin{equation}
f_{gas}=10^{B_2}\bigg(\frac{M_{\Delta_c}}{\hMsun}\bigg)^{A2}\label{eqMf}
\end{equation}
It can be easily deduced  that introducing   the above  mass dependence  $f_{gas}$ into  the $Y-M$ scaling relation  is equivalent  to splitting  Eq.  \ref{eqYM} into two expressions described by Eq.(\ref{eqYf}) and Eq.(\ref{eqMf}), whose best-fit parameters are connected by the simple relation:
\begin{eqnarray}
A = A_1 + A_2 \label{A12}\\
B = B_1 + B_2 \label{B12}
\end{eqnarray}

\begin{figure*}
\centering\includegraphics[angle=0,width=18cm]{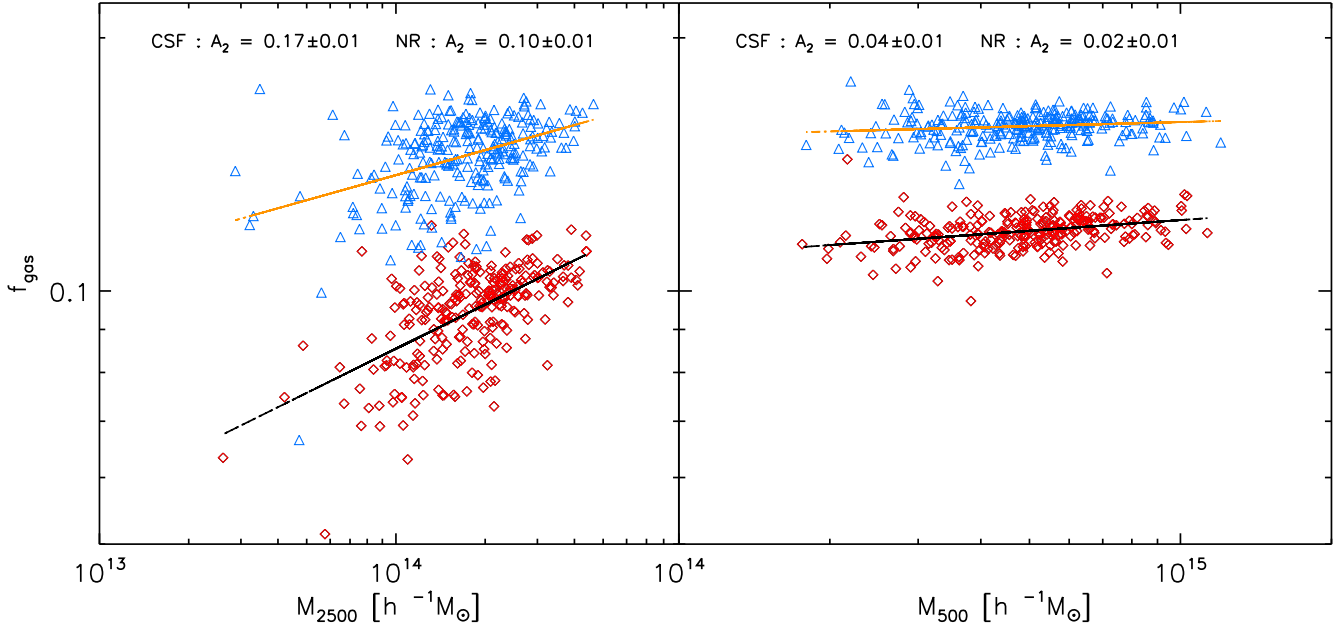}
\caption{The $f_{gas}-M$ scaling relation at $z$ = 0 for $\Delta_c$=2500 (left panels) and $\Delta_c$ = 500 (right panels). CSF clusters are shown in red diamonds, NR clusters in blue triangles. The values of the slope $A_2$ are also reported in the label.}
\label{Mfgas}
\end{figure*}

Since we are  dealing with  massive clusters,   we expect a
 linear relation (in the log space) between the gas fraction and the total 
 mass \citep{VI09}.    In  Fig. \ref{Mfgas} we represent the   results for the $f_{gas}-M$ from our MUSIC clusters  for two  different overdensities.   As can be seen in this figure and in  Table   \ref{tabMF},    the relation is steeper  at the central parts of the clusters and it is shallower in the outskirts while the scatter also increases towards the centre.   To test for the  effects of cooling and star formation in this and other scaling relations, we always compare them with the results from the non radiative simulations of the same clusters.   We see from Fig .\ref{Mfgas} and Table  \ref{tabMF})    that a similar dependence on mass is also present   in NR clusters, although somewhat shallower than in the case of CSF clusters. 

 We can therefore express the best-fit gas fraction
 dependence  on  mass at $\Delta_c$ = 500 as: 
\begin{equation}
f_{gas}=10^{-1.6\pm0.1}\bigg(\frac{M_{500}}{\hMsun}\bigg)^{0.04\pm0.01}
\end{equation}
As we said, this relation becomes much less steep at high overdensities, where  the $f_{gas}-M$ scaling relation at $\Delta_c$ = 2500 can be written as:
\begin{equation}
f_{gas}=10^{-3.5\pm0.2}\bigg(\frac{M_{2500}}{\hMsun}\bigg)^{0.17\pm0.01}
\end{equation}

\begin{figure}
\centering\includegraphics[angle=0,width=8.5cm]{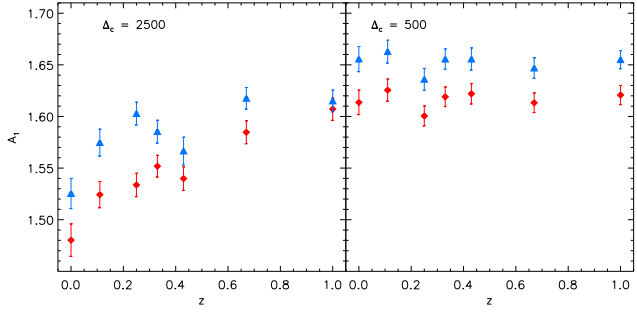}
\caption{Evolution of the A$_1$ slope parameter of the $Yf_{gas}^{-1}-M$ scaling relation at $\Delta_c$ = 2500 (left panel ) and $\Delta_c$=500 (right panel) for CSF clusters (red diamonds) and NR clusters (blue triangles).}
\label{sec6b}
\end{figure}

\begin{figure}
\centering\includegraphics[angle=0,width=9cm]{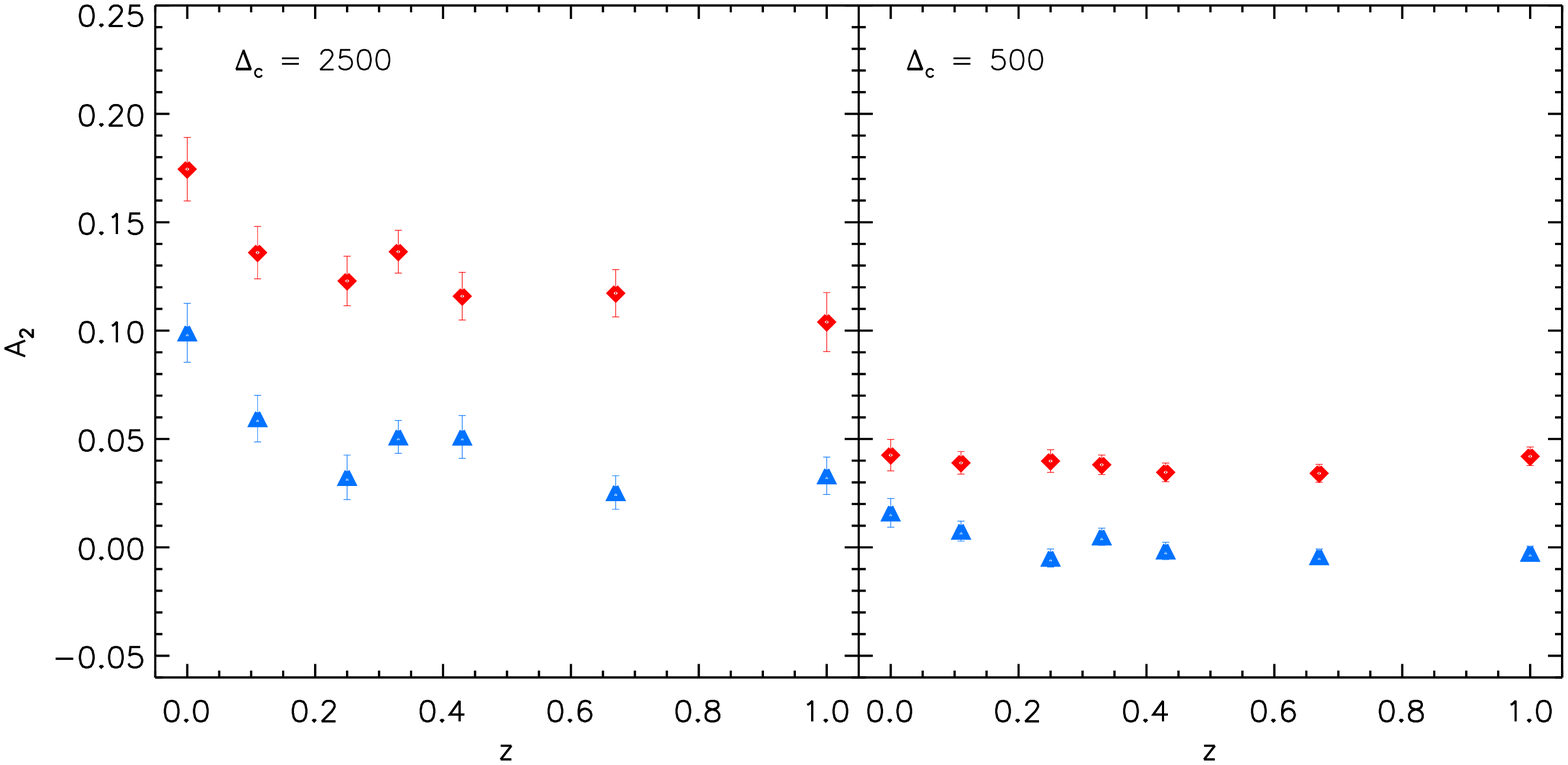}
\caption{ Same as  Fig \ref{sec6b} but for the $A_2 $ slope. }
\label{sec6c}
\end{figure}

\begin{figure}
\centering\includegraphics[angle=0,width=9cm]{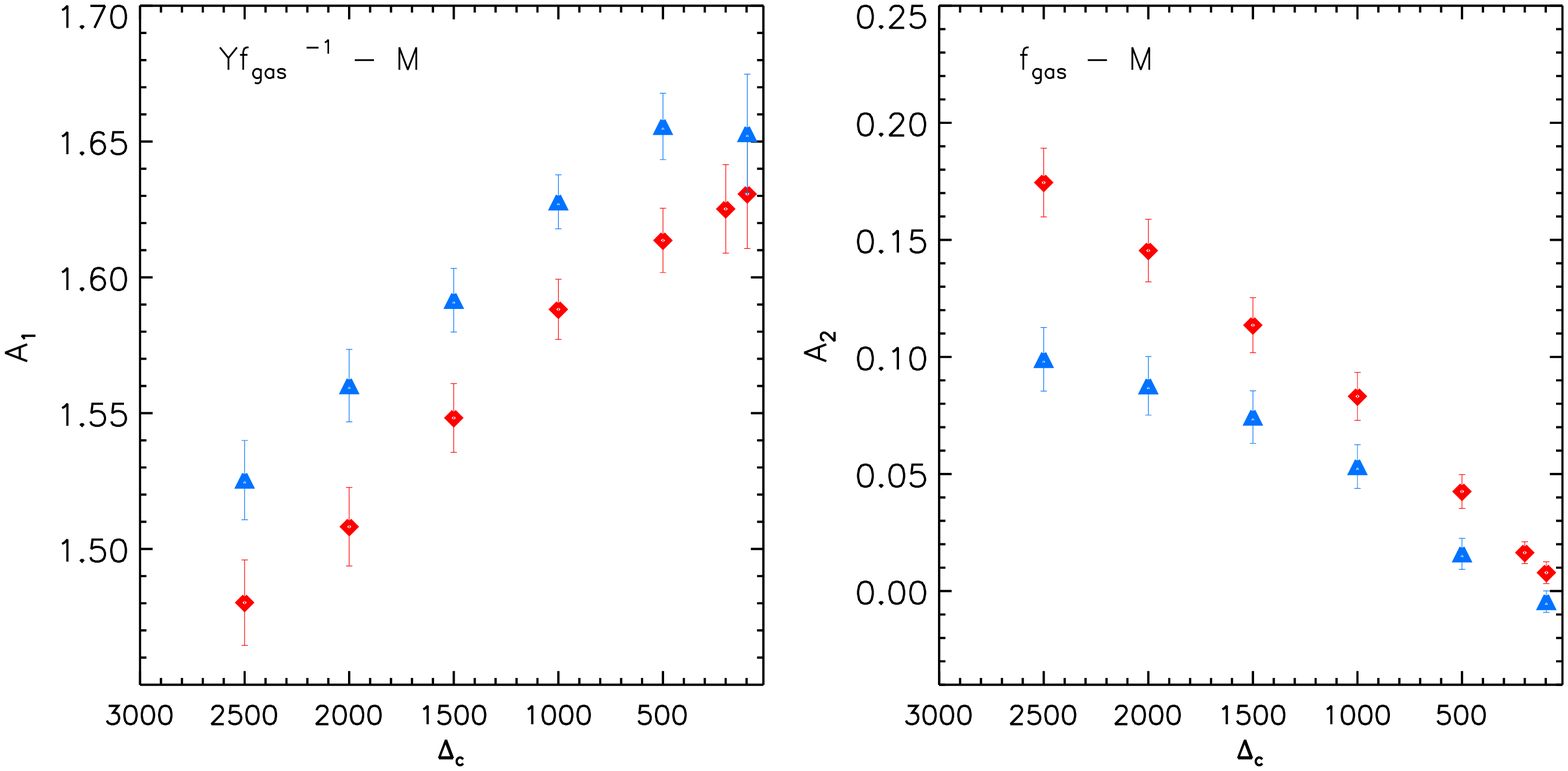}
\caption{ Left panel: the  $A_1$ slope as a function of overdensity at $z=0$  for CSF clusters (red diamonds) and NR clusters (blue triangles).
 Right panel:  Same as left panel but for the $A_2$ slope. }
\label{sec6a}
\end{figure}

In Figs. \ref{sec6b} and \ref{sec6c} we represent the  evolution with redshift of the  average slopes, 
$A_1$ and $A_2$,  for  CFS and NR  clusters  at two  overdensities.   In Fig.   \ref{sec6a} we also show the dependence of  the same  slopes on overdensity for $z=0$   As can be seen, there is a clear linear relation of $ A_1$ and $A_2$  on $\Delta_c$. We can therefore parametrize the variation of the slopes of the $Yf_{gas}^{-1} -M$ and $f_{gas}-M$ scaling relations expressed in equations \ref{eqYf} and \ref{eqMf} as follows:
\begin{equation}
A_1=1.64-0.65\times\bigg(\frac{\Delta_c}{10^4}\bigg)
\end{equation}
\begin{equation}
A_2=0.002-0.7\times\bigg(\frac{\Delta_c}{10^4}\bigg)
\end{equation}

\begin{table*}
\begin{center}
\begin{tabular}{|c|c|c|c|c|c|c|c|c|}
\hline
 & \multicolumn{2}{|c|}{\textbf{$\Delta_c$ (CSF)}} & \multicolumn{2}{|c|}{\textbf{$\Delta_b(z)$ (CSF)}} & \multicolumn{2}{|c|}{\textbf{$\Delta_c$ (NR)}} & \multicolumn{2}{|c|}{\textbf{$\Delta_b(z)$ (NR)}}  \\
 \hline
 & 500 & 2500 & 1500($z$) & 7000($z$) &500 & 2500 & 1500($z$) & 7000($z$) \\
  \hline
\textbf{$\alpha_A$}(10$^{-2}$) & 0.01$\pm$0.90 & 3.4$\pm$1.4 & 0.09$\pm$0.80 & 5.3$\pm$1.4 & -1.7$\pm$1.4 & 1.6$\pm$1.0 & -1.4$\pm$1.3 & -1.6$\pm$0.8 \\
 \hline
\textbf{$\alpha_B$}(10$^{-2}$) & 0.07$\pm$0.80 & 2.7$\pm$1.1 & 1.3$\pm$0.7 & 4.5$\pm$1.2 & -1.3$\pm$1.2 & 1.5$\pm$1.0 & -0.3$\pm$1.1 & -0.6$\pm$0.6 \\
\hline
\end{tabular}
\end{center}
\caption{Best fit parameters $\alpha_A$, $\alpha_B$ of the evolution of $A$ and
  $B$ with redshift at different overdensities and physical proccesses. }
\label{zev}
\end{table*}

Moreover,  from  the values of the best fit parameters listed in Table  \ref{tabMF},    the relation of the slopes  for the different scaling relations expressed   by Eq.(\ref{A12}) and Eq.(\ref{B12}), is confirmed by  our clusters.

\section{Redshift evolution of  the  $Y-M$ scaling relations } \label{sec:results}
In the previous sections we have focused on the study of the $Y-M$ scaling relations from  the MUSIC clusters at $z=0$.  
In this section we will study the possible dependence of the SZ scaling relations with redshift. 

We will employ two distinct approaches in our study: 
\begin{itemize}
\item    Selecting different complete volume limited samples  above a given mass threshold for different redshift and  repeating the study   done at $z=0$.  This will give us the variations of the  $A$ and $B$  best fit parameters of the $Y-M$ relation with redshift (\citealt{SHORT10}; \citealt{FABJAN11}; \citealt{KAY12}).  Following previous studies we can parametrize the redshift dependence  as: 
\begin{eqnarray}
&& A(z) = A_0(1+z)^{\alpha_A}\\
&& B(z) = B _0(1+z)^{\alpha_B}
\end{eqnarray}
where $A_0$ and $B_0$ are the values of the slope at $z$ = 0 and
$\alpha_{A,B}$ the best fit  parameters that will describe the  possible dependence with
redshift.

\item An  alternative approach, more similar to observational studies,  consisting of  generating a sample of clusters   that are carefully selected from  different redshifts, by  keeping the same proportion   of abundance of clusters above a given mass at each redshift considered.  From these samples, we look for possible dependence on redshift of the form:
\begin{equation}
Yf_{gas}^{-1}E(z)^{-2/3} =  BM^A(1+z)^{\beta}\label{ym_z}
\end{equation}
A similar approach has been adopted in \cite{FERR11} with 438 clusters
of the X-ray Galaxy Cluster Database (BAX) having redshifts ranging between 0.003 and 1 to obtain a large sample of potentially observable
clusters in SZ.
\end{itemize}

Since we want to study the redshift evolution of the scaling relations
for the  two different definitions of aperture radius based on the 
overdensity criteria  introduced in section \ref{sec:bar}  (and discussed
in  more detail in Appendix A), we have to make some preliminary considerations.
When we use the redshift dependent background overdensity to define the
integration area,  we have to replace  $\Delta_c$  in Eq.  (\ref{YS})
of the    $Y-M$ relation  by:
\beq
\Delta_c=\Delta_b\cdot\bigg(\frac{\Delta_v(z)}{\Delta_v(0)}\bigg)\times\Omega_m(z).
\eeq
In this case, the normalization parameter $B$ is transformed as:
\begin{equation}
B = \log \frac{\sigma _T}{m_ec^2}\frac{\mu}{\mu _e}\bigg(\frac{\sqrt{\Delta_c}GH_0}{4}\bigg)^{2/3} + \log f_{gas}\bigg(\frac{\Delta_{b,v}(z)}{\Delta_{b,v}(0)}\Omega_m(z)\bigg)^{1/3}
\end{equation}
But this introduces a natural evolution with redshift  due to the
scaling of $\Delta_b(z)$ that has to be subtracted from the fits to
properly measure any intrinsic evolution of  $A$ and $B$.

\subsection{Evolution with redshift of the $Y-M$ best fit parameters}

\begin{figure*}
\centering\includegraphics[angle=0,width=14cm]{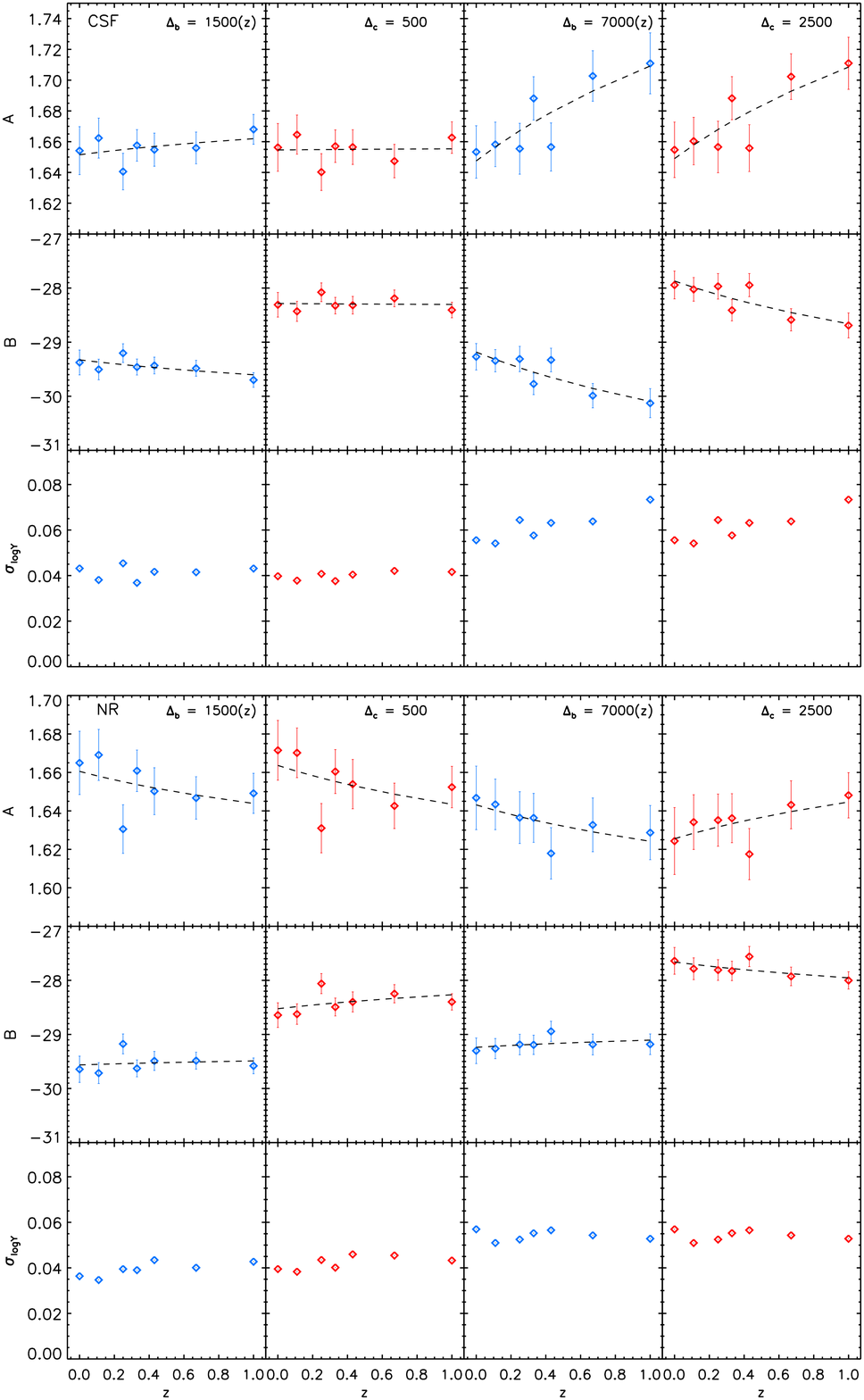}
\caption{Evolution with redshift of the slope (top panels), of the normalization (central panels) and of the scatter (bottom panels). The upper part (first three rows) of the figure describe CSF clusters, the lower (last three rows) the NR clusters. Redshift varying background density (blue diamonds) at $\Delta_b$=1500($z$) (left) and $\Delta_b$=7000($z$) (right) and fixed critical overdensity (red diamonds) at $\Delta_c$=500 (left) and $\Delta_c$=2500 (right).}
\label{CSF}
\end{figure*}

In Fig.\ref{CSF} we represent the mean values of the best fit   $A$ and
$B$  parameters of the complete volume limited  mass samples  for CSF
clusters (upper panels) at the same  four different overdensities already
analyzed in section \ref{sec:bar}. We also estimated  the scatter on
$\sigma_{\log Y}$ for the same cases,  calculated as in Eq.(\ref{sigma}). 

We can appreciate that the $A$ slope and the normalization $B$  show
a clear evolution with redshift for low overdensities while they flat
off when higher overdensity values are considered.   The situation
is the opposite  when we consider higher overdensities: at  these apertures the
slope seems to be flatten  off up  to redshift $z\sim$ 0.5 and then
increases. 

Table  \ref{zev} shows the value of the best fit parameters
$\alpha_A$ and $\alpha_B$  for  the same overdensities analyzed in the previous
sections. The results support the hypothesis of  null evolution of $Y-M$ scaling relations at
small  overdensities ($\Delta_c$=500, $\Delta_b$=1500($z$)) and a possible mild evolution
at high overdensities ($\Delta_c$=2500, $\Delta_b$=7000($z$)). 

 The total mass  of the cluster seems to play a fundamental role in the
evolution of the scaling relations.  The same is also certain for  the $f_{gas}-M$
scaling relation. The $Y-M$ relation seems therefore to satisfy
self-similarity when considering massive evolved objects (we recall
that all the clusters analyzed in this paper have $M_v >$
5$\times$10$^{14}h^{-1}$ at $z$=0). When smaller mass systems are added to the
catalogues, the deviations from  self-similarity  become more apparent,
as it is clearly shown in Table.  \ref{zev}

The rms scatter of $\log_{10} Y$, $\sigma _{\log Y}$, is also shown in  the
lower row of the upper  panel of  Fig.\ref{CSF}. Although it is  nearly
constant at low overdensities (about
4 per cent), its value increases (from 6 per cent to 8 per cent) at high overdensities
and high redshifts.

If we look at  the same evolution in NR clusters (bottom panels of Fig.\ref{CSF}), we
find that,  in this case,  there is no evidence of any change in the slope
with redshift, and the values of $\alpha_A$ and $\alpha_B$ are extremely low at
all overdensities (Table  \ref{zev}). Also the scatter  is  small (never
exceeding 4 per cent) and does   not vary with redshift. 

$\;$  From Fig.\ref{CSF} and Table  \ref{zev} we can also
conclude that  the evolution of $A$ and $B$ is  not affected by the
choice of  the overdensity (fixed critical or redshift dependent 
background), since  both methods lead to the same results (evolution at
high overdensities and smaller masses for CSF clusters and  no
significant evolution for NR clusters). Therefore,  there is no need to
change  from the widely used, and more simple,   assumption of a fixed critical overdensity 
to define the aperture radius.

\begin{table}
\begin{center}
\begin{tabular}{|c|c|c|c|c|c|c|c|}
\hline
\textbf{z} & 0.00 & 0.11 & 0.25 & 0.33 & 0.43 & 0.67 & 1.00 \\
\hline
\textbf{N} & 271 & 237 & 187 & 147 & 117 & 44 & 8 \\
\hline
\end{tabular}
\end{center}
\caption{Number of MUSIC-2 clusters with M$_v>$5$\times$10$^{14}$\hMsun at different redshifts.}
\label{Ncl}
\end{table}

\subsection{Mixed-z sample results}

\begin{figure*}
 \centering
 \includegraphics [width=\textwidth]{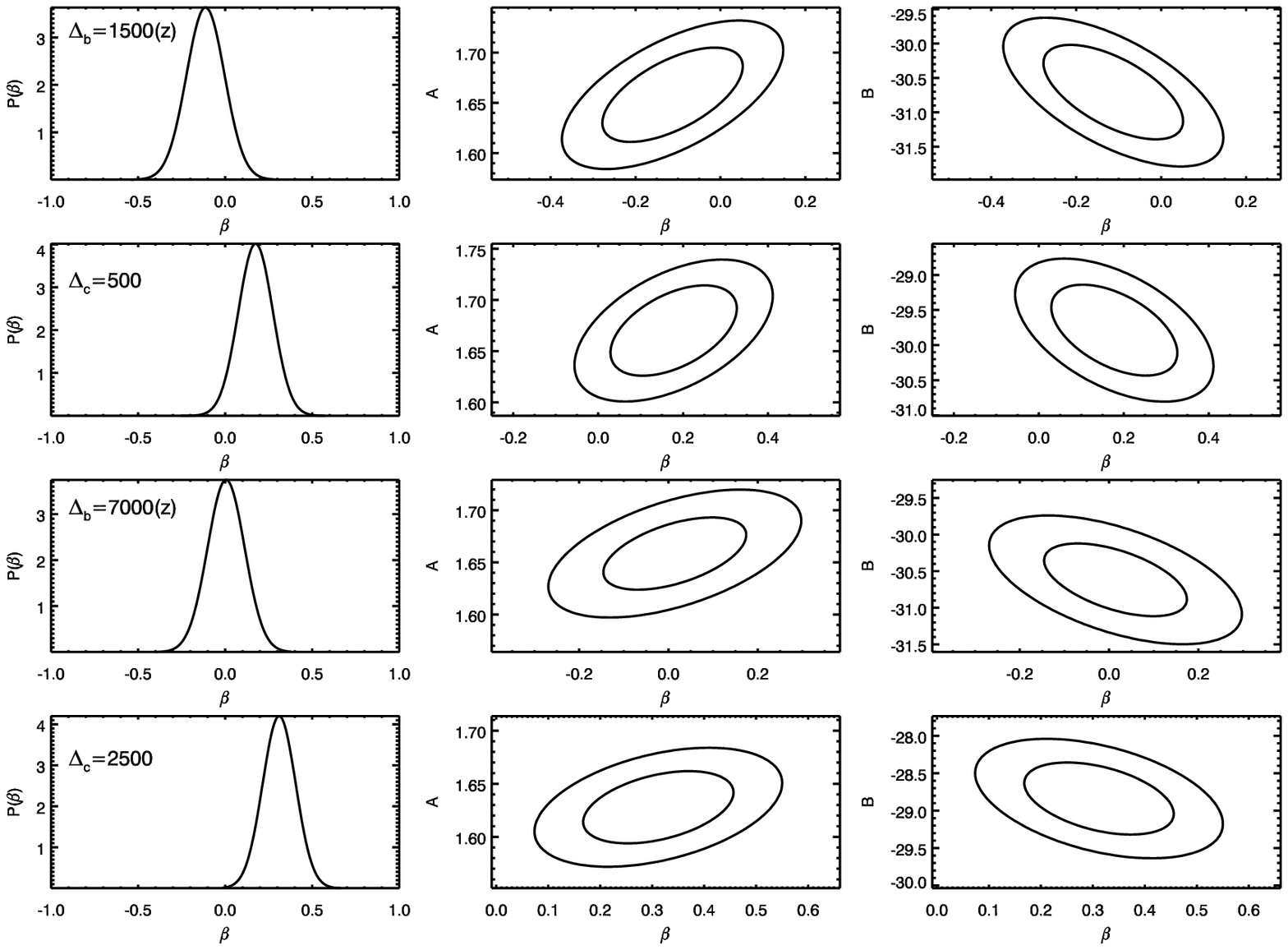}
\caption{Posterior distributions for the $\beta$ parameter at the 4 overdensities considered: $\Delta_b$=1500($z$), $\Delta_c$=500, $\Delta_b$=7000($z$), $\Delta_c$=2500. The contours correspond to 1 and 2 $\sigma$ intervals for the best fit parameters.}\label{fig:threeparamposteriors}
 \end{figure*}
 
As we said above, we have also checked the evolution of the $Y-M$ scaling relation on
mixed-$z$ datasets to mimic the variety of clusters that should
populate an observational  mass-limited galaxy cluster survey. These
datasets have been  built by
selecting subsamples of CSF objects from all the available redshifts,
with the prescription  that every member appears in the sample only at
one redshift and that the sample is populated according to the cluster
abundances of  MUSIC-2 as a function of redshift (see Table
\ref{Ncl}).  This approach allows us to evaluate systematics in the
reconstruction of the $Y-M$ scaling relation. Moreover,  if some evolution or
mass dependence actually exists, it will also help us to estimate the
level at which it is detectable  in a complete mass-limited cluster sample

\begin{figure}
 \centering
 \includegraphics [width=9cm]{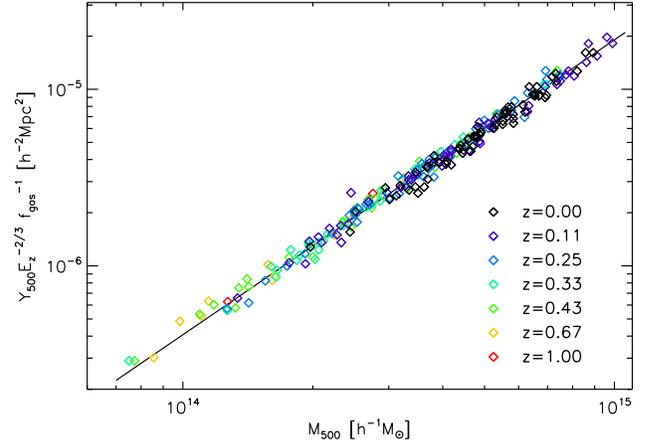}
\caption{Mixed-$z$ scaling relation at $\Delta_c$ = 500; the different diamond colors refer to different redshifts between $z$ = 0 and $z$ = 1.}
\label{zmix500}
 \end{figure}

\begin{figure}
 \centering
 \includegraphics [width=9cm]{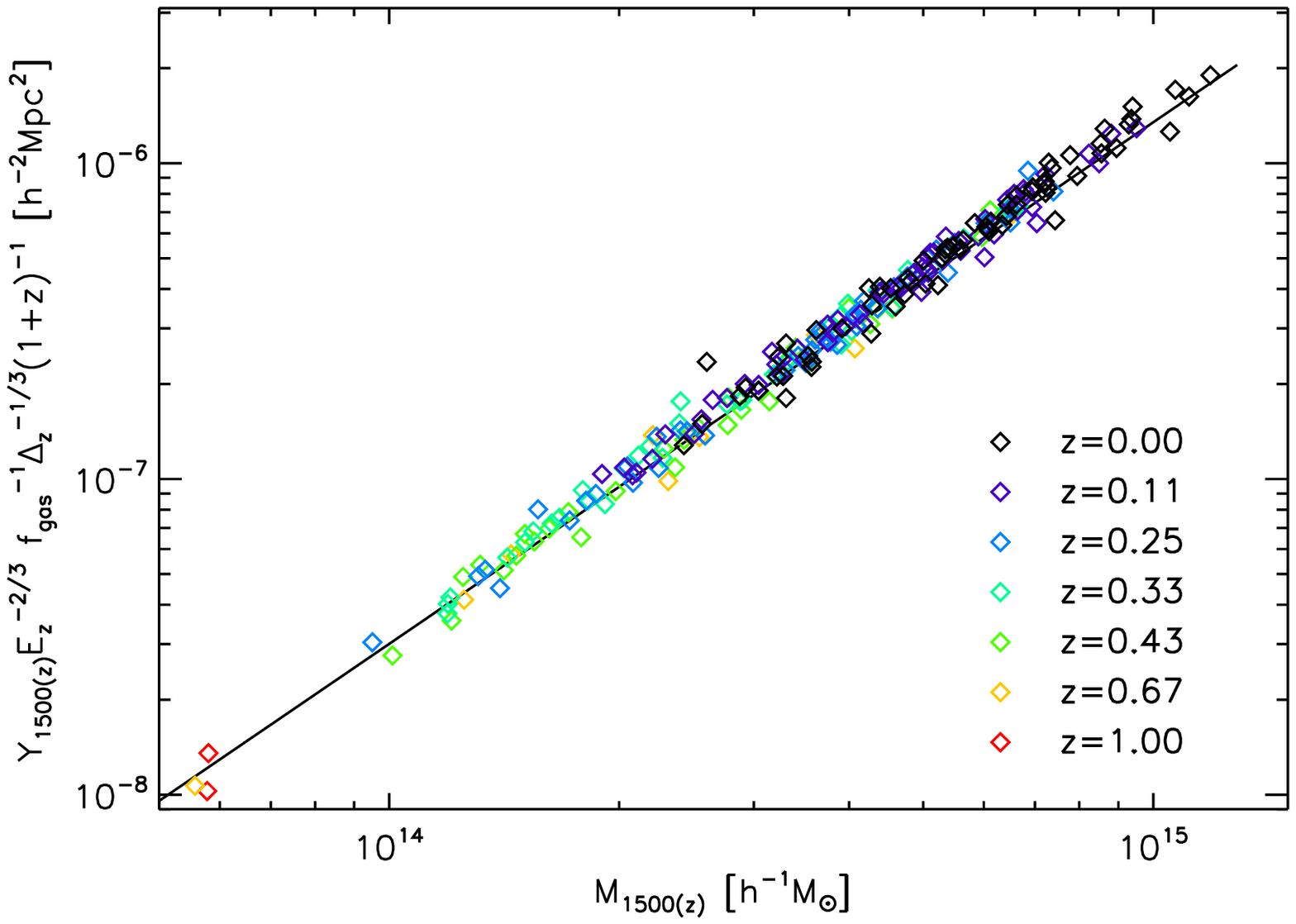}
\caption{Mixed-$z$ scaling relation at $\Delta_b$ = 1500($z$); the different diamond colors refer to different redshifts between $z$ = 0 and $z$ = 1.}
\label{zmix1500}
 \end{figure}

 We assume Eq.(\ref{ym_z}) as our
reference scaling law: 
\beq
\log (YE(z)^{-2/3}f_{gas}^{-1})=B+A\log M + \beta \log(1+z)
\eeq
The fit has been performed with a standard Monte Carlo Markov Chain
algorithm based on the Metropolis-Hastings sampling scheme, where no
additional priors have been applied on the slope parameter. The
convergence has been assessed through the Gelman-Rubin test
\citep{GEL92} on a set of three simultaneous chains run on the same
dataset from independent starting points in parameter space.  
The samples were drawn at  the four reference overdensities (again $\Delta_c$ = 2500, 500 and $
\Delta_b(z)$=7000($z$), 1500($z$)). Fig. \ref{fig:threeparamposteriors}
shows the posterior distributions of $\beta$ and the $\beta - A$ and $\beta - B$
joint likelihoods for the  reference overdensities. The fit results
are still fairly consistent with self similarity and no additional
redshift scaling in the $Y-M$ scaling relation, except for the
$\Delta_c=2500$ dataset, which shows marginal evidence of residual scaling. 
Since the mixed-$z$ cluster sample has been drawn from the original
simulated dataset randomly, the procedure was iterated on 5000 samples
to check for  the possibility of large fluctuations in the fit results
(e.g. due to outliers in the data distribution). This was not the case,
since the best-fit parameters at the 4 overdensities lied well within
the range of fluctuation allowed by the individual-sample posterior
distributions.  In Table   \ref{three_params} all the 3 best fit
parameters are listed with their corresponding 1 $\sigma$ errors. 
The $Y-M$ scaling laws for one realization of a mixed-$z$ sample and
only for low overdensities, are plotted  in Figs. \ref{zmix500} and
\ref{zmix1500}.

\begin{table}
\centering
\begin{tabular}{ c || c | c | c | c }
{}           & $\Delta_c$ =500 & $\Delta_c$ =2500 & $\Delta_b$ = 1500(z) & $\Delta_b$ = 7000(z) \\
\hline
A &1.672$\pm$ 0.028&1.627 $\pm$ 0.022&1.652$\pm$ 0.028&1.656 $\pm$ 0.023\\
B &-29.77$\pm$ 0.42&-28.85$\pm$ 0.32&-30.65$\pm$0.42&-30.61$\pm$ 0.33\\
$\beta$ &0.17 $\pm$ 0.10&0.31$\pm$ 0.14&-0.12$\pm$ 0.11&0.00$\pm$ 0.11\\
\end{tabular}
\caption{Best fit parameters (slope A, normalization B and redshift evolution $\beta$) for the $Y$-$M$ scaling relation on a mixed-$z$ sample of 250 CSF clusters. }
\label{three_params}
\end{table}


\section{Summary and Conclusions} \label{sec:conclusions}

Recent new large scale surveys on the far infrared and millimeter bands
confirmed the validity of the Sunyaev-Zel'dovich Effect as a useful
cosmological tool to detect and to study galaxy clusters. Today,  one of
its most relevant applications is to infer the total cluster mass using   the $Y-M$ scaling relation.  
Recent observations (Planck, SPT and ACT)  provide the SZ-measurements of a large number of clusters from which the 
$Y-M$ relation can be reliably   inferred.  

In this paper we  have presented  the MUSIC  dataset,  a large sample of gasdynamical resimulated
massive clusters, that have been selected from the  MareNostrum Universe simulation and from the  MultiDark simulation. 
Our dataset comprises two  different subsamples. The MUSIC-1  which contains  164 objects  that have been selected according to their 
  dynamical state  (i.e. bullet-like clusters and relaxed clusters) and the MUSIC-2  which contains  a total of more than 2000   galaxy groups and clusters.  The MUSIC-2 constitutes a   complete volume limited sample for clusters with virial masses above   $8\times 10^{14} h^{-1} M_\odot$ at $z=0$.   To our knowledge,  there are no other works about resimulated clusters that  have shown such a large number of massive objects.  These objects have been simulated using two physical   processes:   non-radiative (NR, gravitational heating only)  and a radiative model (CSF) including  cooling, UV photoionisation,  multiphase ISM and star formation and supernovae feedbacks.  No AGN feedback has been included in the radiative simulations. 

We  plan to  generate mock catalogues   from the MUSIC clusters  for the  most common  observables  such as: X-ray  and SZE surface brightness maps,  lensing maps and  galaxy  luminosity functions.  All these deliverables, together with the initial conditions for all the MUSIC simulations, will be made publicly available from the website http://music.ft.uam.es. 

Here we have presented the first results from MUSIC  based on the analysis  of the baryon content  and   the scaling relations of the  thermal SZ effect. 

The analysis of the integrated cluster quantities, such as the fractions  of the different baryon components, or $Y$ parameter, have been  done using two definitions of the aperture radius.  The most commonly used  approach assumes  that the integration domain is defined by a fixed value of the  mean overdensity inside the cluster region compared to the critical overdensity of the Universe.   It has been shown (see Appendix A)  that this definition  introduces a bias  when  any  integrated  quantity    is compared  at different redshifts.   On the contrary, by using a redshift-dependent overdensity compared to the background density of the Universe we can rescale the aperture radius  such that  the integrated region is always the same volume fraction of the total cluster volume defined by the virial radius.

Our numerical results  on  the baryon content and the SZ scaling relations have  been explored to check possible dependences on  overdensity, redshift and cluster physics modelling.  We have also   made a comparison  between  MUSIC clusters  and available  observational results.  
The scaling relation  between SZ brightness, quantified with the integrated Compton $y$-parameter, $Y$, and the cluster total  mass, $M$, confirmed  the robustness of the self-similarity assumption for this class of objects present in MUSIC sample.  We evaluated  the effect  on the $Y-M$  of having    the cluster gas fraction,  $f_{gas}$,   as  a free parameter that can vary with the cluster mass.   Finally, we  looked for a possible  redshift evolution of the $Y-M$ relation  using two  different approaches. 

Our main conclusions can be summarized as follows:

\begin{itemize}
\item The mean value of the gas fraction of   MUSIC-2  CSF clusters  at $\Delta_c$=500 is $f_{gas}$=0.118$\pm$0.005.  This value is  compatible with observational results shown in \cite{MGAN04}, \cite{LAROQUE06}, \cite{MGAN06},  \cite{VI09}, \cite{ZHANG10}, \cite{ET10}, \cite{JU10}, and \cite{DJF12}. The values of the baryon fraction at virial radius, for both simulation flavours  (CSF and NR), are consistent,  within the errors,  with the cosmological ratio $\Omega_b/\Omega_m$ (according to WMAP7 cosmology). At higher overdensities (i.e. $\Delta_c$=2500),  we also find agreement for the $f_{gas}$  with observational results from \cite{MGAN06}, \cite{VI09} and  \cite{ZHANG10} and  we are marginally consistent with  \cite{AL08} and \cite{DJF12}.  This means that the possible effects of numerical overcooling  do  not appear to affect dramatically  our simulations, despite the fact that  we  have not included strong  AGN feedback  in our model. \textbf{The star fraction measured in MUSIC-2 CSF clusters does not agree with the one estimated by observations, which for massive clusters predict a star fraction considerably smaller than one extracted from MUSIC simulations.}
Concerning the evolution with redshift of the  different baryon components, we  find  significant differences in the evolution depending on the definition of aperture radius   used.  This is more evident at large overdensities, where the considered fraction of virial radius increases with redshift
when a fixed  critical overdensity is used  to the define it (see Fig. \ref{rad}).

\item The $Y-M$ scaling relation  derived from our  clusters, assuming a fixed $f_{gas}$,  agrees very well with the predictions of the self-similar  model and shows a very  low dispersion ($\sigma_{logY}$ $\simeq$ 0.04).  The resulting  best fit $Y_{500}-M_{500}$ relation can  then be expressed as:
\begin{equation}
Y_{500} = 10^{-28.3\pm 0.2}\bigg(\frac{M_{500}}{\hMsun}\bigg)^{1.66\pm 0.02}E(z)^{-2/3}[h^{-2}Mpc^{2}]
\end{equation}
The $M-Y$ relation,  which is more suitable to infer cluster masses  from $Y$ measurements, is also compatible with the self-similar model
 (which predicts a slope of 3/5) with an even lower dispersion ($\sigma_{logM}\simeq$0.03). The corresponding  best-fit $M_{500}-Y_{500}$ relation can be expressed as:
\begin{equation}
M_{500} = 10^{17.0\pm 0.1}\bigg(\frac{Y_{500}}{h^{-2}Mpc^2}\bigg)^{0.59\pm 0.01}E(z)^{-2/5}[\hMsun]
\end{equation}
We have compared the above fits with the recent results on the $Y-M$  from the  Planck  Collaboration.    The  agreement  is very good, provided that the masses of Planck clusters are overestimated  by   22 per cent  due to the biases   of the  X-ray mass estimations with respect to the lensing measurements.      We have tested whether this bias could be due to the  lack of hydrostatic equilibrium  hypothesis. By comparing  the hydrostatic masses with the true one in our MUSIC samples,  we find  a negative  HMB (an opposite trend  than in Planck clusters).  In our case, we find  25 per cent underestimation  of the true mass    by the hydrostatic mass, \textbf{in agreement with other simulations studying the same topic which already found that the HSE hypothesis introduce a negative bias on the true mass estimation.}

\item The dependence of  the gas fraction on  the cluster total mass has been also studied along the  cluster aperture  radius.  There is a linear relation between $f_{gas}$  and total mass which is more evident, although with  larger scatter, when we approach the cluster core. 
This effect  is rather  insensitive  on the adopted physics in the numerical simulations: both CSF and NR clusters present a similar behavior.

\item    Leaving $f_{gas}$ as a free parameter in the $Y-M$ scaling relation leads to a deviation from self-similarity.   The  modified $Yf_{gas}^{-1} -M $  relation  and the $f_{gas}-M$ relation are directly related  with  the $Y-M$   if both of them are  expressed as power laws. We confirm this hypothesis and provide the fits to the  $f_{gas} -M $  for  two overdensities $\Delta_c=2500$ and $\Delta_c=500$
\begin{equation}
f_{gas}=10^{-3.5\pm0.2}\bigg(\frac{M_{2500}}{\hMsun}\bigg)^{0.17\pm0.01}
\end{equation}
\begin{equation}
f_{gas}=10^{-1.6\pm0.1}\bigg(\frac{M_{500}}{\hMsun}\bigg)^{0.04\pm0.01}
\end{equation}

\item We have studied the redshift  evolution of the slope, and of the normalization, in the $Y-M$ scaling law.  We did not find any evolution for NR clusters. 
On the contrary,  CSF clusters show a marginal deviation of the slope from the self-similar prediction at high overdensity for redshifts larger than 0.5. This result is confirmed   using two different methods to study the evolution with redshift.

\item No significant differences  have been  found  when comparing results  at different redshifts using the two  alternative definitions of aperture areas based on  different overdensity criteria:   the standard, redshift independent, critical overdensity and the redshift-dependent background overdensity show the same results, at least for the gas fraction analysis, with the latter appearing to be more sensitive to the redshift evolution of the sample properties ($Y-M$ slope, gas fraction, etc.). We can therefore conclude that even if the redshift-dependent background overdensity leads to results which better identify a possible evolution, no significant errors are introduced  in the analysis of the SZ effects   when  using the standard critical overdensity criteria.  The same conclusions  in the case of X-ray   analyzes have been raised by   \cite{MGAN06} .

\item  The use of radiative physics on galaxy cluster simulations introduces fundamental improvements respect to non-radiative simulations: CSF clusters show better agreement with observations in the estimate of the gas fraction  as well as in the study of the $Y-M$ scaling relation. On the other hand, NR clusters overestimate the gas fraction and do not seem  to be compatible with $Y-M$ scaling relations from observational results.

\end{itemize}

In summary, we have shown that the MUSIC dataset is well suited for the study of massive cluster properties and provides a reasonable description of observed objects with similar mass. Therefore, the MUSIC clusters can be a good  cosmological tool.  In upcoming papers we are going to extend the analysis to  other complementary scaling relations in X-ray, lensing and optical in order to have a full set of  observables for  a complete volume limited sample of  $\Lambda$CDM simulated galaxy clusters.

\appendix
\section{Cluster aperture radius definitions }

A  correct analysis of the  integrated properties of   clusters at similar  or different redshift,  is based on the 
definition of  the  aperture radius. Ideally the virial radius is a suitable choice, confining almost the whole cluster.
Unfortunately a such large region is challenging to be explored due to the lack of sensitivity of present day  SZ or X-ray observations. 
Therefore, a smaller aperture radius has to be defined.    This quantity has to be physically meaningful  and  must  hold  at different redshifts.

The widely applied definition of  the aperture radius, both in observations and simulations,   correspond to the radius of a sphere  that contains a total mass that equals the mass of a sphere with a  fixed redshift independent overdensity  with respect to the critical density defined as 
 \beq
\rho_c(z)=\frac{3H_0^2E(z)^2}{8\pi G}
\eeq
where {\itshape H$_0$} is the present value of the Hubble constant, G is the universal gravitational constant.

An alternative  definition of the aperture radius consists in defining it as  the radius of the  sphere that contains a total mass equal to the mass of  a sphere with  an overdensity   value with respect to the cosmic background mass density, $\rho_b(z)$ = $\rho_c(z) \Omega_m(z)$  where $\Omega_m(z)$ is the cosmological  matter density parameter defined as $\Omega_m(z) = \Omega_M (1+z)^3 / E^2(z)$. 

The cluster centric radius can then be  quantified in terms of these  two overdensities $\Delta_{b,c}$ , which means that the cluster average density inside \emph{r$_{\Delta_{b,c}}$} is $\Delta$ times the critical or background density. 
The overdensity radius, linked to the critical (or background) density, can therefore be expressed as:
\begin{eqnarray}
r_{\Delta_c} = \bigg({\frac{3M_{\Delta_c}(z)}{4\pi\rho_c(z)\Delta_c(z)}}\bigg)^{1/3}    \label{radius}\\
r_{\Delta_b} = \bigg({\frac{3M_{\Delta_b}(z)\Omega_m(z)}{4\pi\rho_c(z)\Delta_b(z)}}\bigg)^{1/3}    \label{radius_b}
\end{eqnarray}

It is  important to check whether  a redshift independent $\Delta_c$ could possibly introduce some biases  when comparing  clusters at different redshifts.
Our goal is to keep the ratio $r/R_v$ constant with redshift  so that  the  same fraction of cluster is considered at different redshifts.
 
\cite{MGAN06} were the first to address this problem. They investigated  the influence of the definition of overdensity radius only in the case of X-ray scaling relations 
with a sample of 11 high-redshift clusters observed with Chandra and/or XMM-Newton. Also recently a study of scaling laws specifically, for X-ray observables and cluster properties, has faced the same problem \citep{BOHR12}.

The dependence  of  the overdensity radius defined in   Eq.\ref{radius}  on redshift and on  cosmological parameters can be easily derived:
\beq
r_{\Delta_c}\propto \rho_c(z)^{-1/3}\propto E^{-2/3}(z)=(\Omega_M(1+z)^3+\Omega_{\Lambda})^{-1/3}\label{r_c}
\eeq

Similarly, the radius defined with the background overdensity depends on redshift as
\beq
r_{\Delta_b} \propto \rho_c(z)^{-1/3}\Omega_m(z)^{-1/3} \propto (1+z)^{-1} \label{bkg}
\eeq

As can be seen, the dependence on redshift of the two defined radius is somewhat different.  From $z=0$ to $z=1$, the $r_{\Delta_c} $ decreases by a factor of  1.4  while the $r_{\Delta_b}$ decreases a factor of 2.  But,  we are interested to check  how the ratio  $r_{\Delta_{c,b}}/R_v$ depends on redshift.  Thus, we take into account the dependence of the overdensity at virial radius $R_v$, $\Delta_v$. An analytical fit  to the numerical solution  of  the spherical collapse model  was given by   \cite {BN1998} in a $\Lambda$CDM cosmology (see also \cite{EKE}). 
\begin{equation}
\Delta_{v,c}(z) \sim 18 \pi^2 + 82 x - 39 x^2 \label{vir2}
\end{equation}
as to the background density, is given by:
\begin{equation}
\Delta_{v,b}(z) \sim \Delta_{v,c}(z)/(1+x) 
\end{equation}
where $x = \Omega_m(z) - 1$.  
 
\begin{figure}
\centering\includegraphics[angle=0,width=10cm]{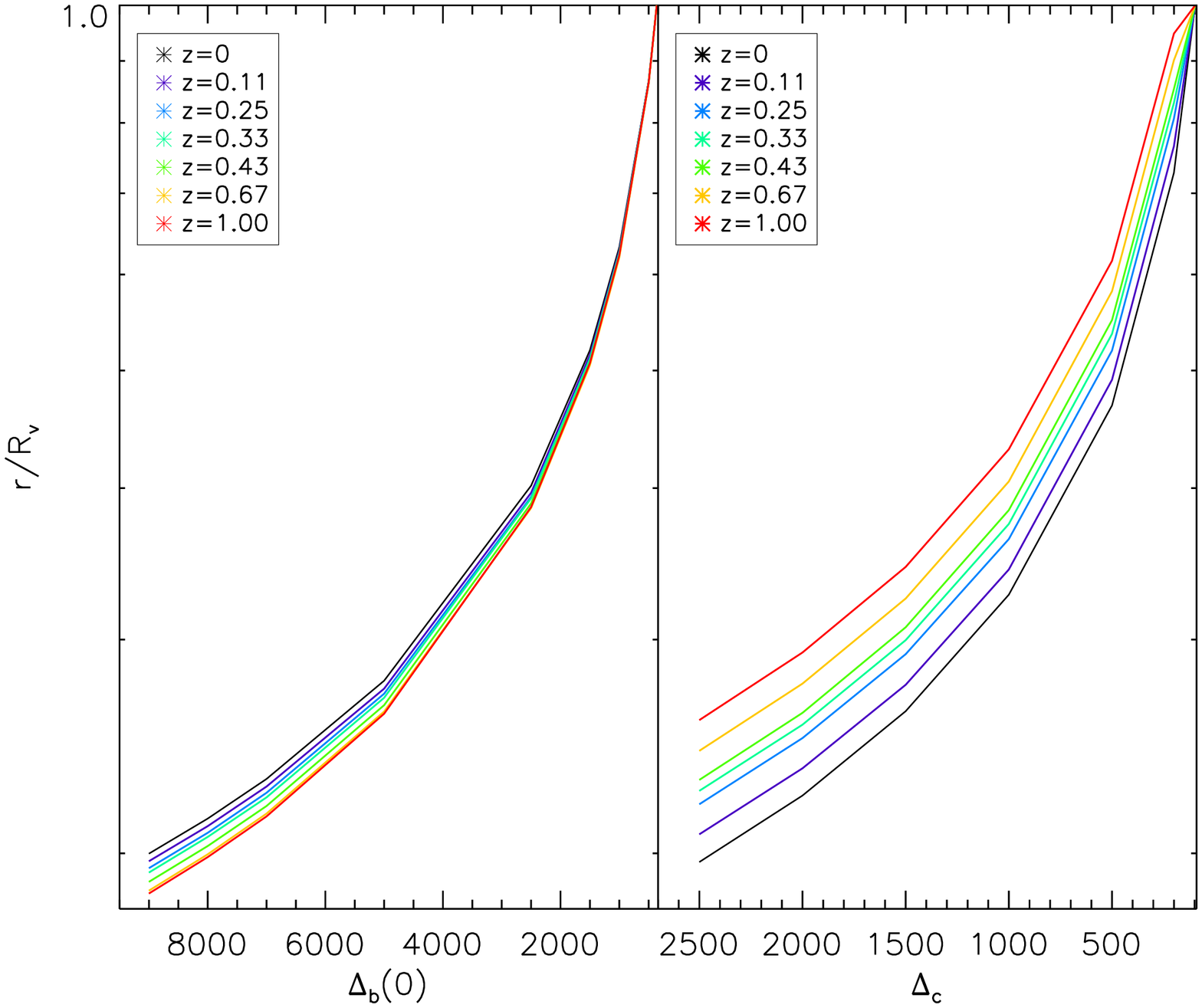}
\caption{Behaviour of the $r/R_v$ ratio at different redshifts (0 $\leq z \leq$ 1) for background overdensity varying with redshift (left panel) and for fixed critical overdensity (right panel).}
\label{rad}
\end{figure}

With this definition of aperture radius the $r_{\Delta_c}/R_v$ ratio  changes with  redshift as it is shown in  right panel of  Fig. \ref{rad}.  The effect increases towards the inner regions of the cluster.  An overdensity radius of 2500 over the critical density  corresponds to a region of $1/5$th of the  virial radius at  $z=0$ and  changes by more than 20 per cent   towards $z=1$.

\cite{MGAN06} suggested  an alternative  definition of  aperture radius  to show an almost redshift independent variation of $r/R_v$.  To this end, they propose to define   the aperture radius with  a  redshift dependent  overdensity  value which is a simple rescaling of the overdensity   defined at $z=0 $ as:

\begin{equation}
\Delta(z) = \Delta(0)\bigg[\frac{\Delta_v(z)}{\Delta_v(0)}\bigg]\label{vir}
\end{equation}

In this way,  the ratio of the comoving densities is ensured to be constant along the redshift. We refer to this redshift varying overdensity as $\Delta(z)$, following the notation adopted by \cite{MGAN06}. 

By looking at the dependence on redshift, cosmology and on the density
contrast at the virial radius, a background overdensity, including the
$z$-dependence (Eq.\ref{vir}), seems to be a more suitable  choice for
the $\Delta$ definition, allowing to compare cluster properties within radii
corresponding to redshift independent cluster fractions. In fact, as can be seen in the left panel of Fig. \ref{rad}   $r_{\Delta_b(z)}/R_v$,  estimated  from the redshift dependent 
background overdensity, $\Delta_b(z)$,  looks  almost invariant with redshift ( less than 5 per cent variation). Thus,  applying this definition,  we will be  able to
study the evolution of cluster properties almost free from the bias  induced
by the choice of an aperture  radius  defined by the same  overdensity  value at all redshifts.

\section*{Acknowledgements}
The  MUSIC simulations
were performed at the Barcelona Supercomputing Center (BSC) and the initial conditions were done at the Leibniz Rechenzentrum
Munich (LRZ). We thank the support of the MICINN Consolider-Ingenio 2010 Programme under grant MultiDark CSD2009-00064. GY acknowledges support from MICINN under research
grants AYA2009-13875-C03-02, FPA2009-08958 and Consolider
Ingenio SyeC CSD2007-0050. 
FS, MDP, LL and BC have been supported by funding from the University of Rome 'Sapienza', {\textbf Anno 2011 - prot. C26A11BYBF}. FS is also supported by the Spanish Ministerio de Ciencia e Innovaci\'on (MICINN) with a FPU fellowship. MDP wish to thank the Ministerio de Educacion, Cultura y Deporte for supporting his visit at UAM during the final phase of preparation of this paper. \textbf{We thank the anonymous referee and Gabriel Pratt for the useful suggestions.}

\bibliographystyle{mn2e}
\bibliography{archive}

\bsp

\label{lastpage}

\end{document}